\documentclass[12pt]{article}
\usepackage{geometry}
\usepackage{a4}
\usepackage{graphicx}
\usepackage{epsf}
\usepackage{amssymb}
\usepackage{cite}
\newcommand{\be}{\begin{equation}}
\newcommand{\ee}{\end{equation}}
\newcommand{\rmnum}[1]{\romannumeral #1}
\newcommand{\Rmnum}[1]{\expandafter\@slowromancap\romannumeral #1@}
\newcommand{\bea}{\begin{eqnarray}}
\newcommand{\eea}{\end{eqnarray}}

\begin{document}
\def\C{{\mathbb{C}}}
\def\R{{\mathbb{R}}}
\def\s{{\mathbb{S}}}
\def\T{{\mathbb{T}}}
\def\Z{{\mathbb{Z}}}
\def\W{{\mathbb{W}}}
\def\Bbb{\mathbb}
\def\BZ{\Bbb Z} \def\BR{\Bbb R}
\def\BW{\Bbb W}
\def\BM{\Bbb M}
\def\BC{\Bbb C} \def\BP{\Bbb P}
\def\CP{\BC\BP}
\begin{titlepage}
\title{On The Phase Structure and Thermodynamic Geometry of R-Charged Black Holes}
\author{}
\date{
Anurag Sahay, Tapobrata Sarkar, Gautam Sengupta
\thanks{\noindent E-mail:~ ashaya, tapo, sengupta @iitk.ac.in}
\vskip0.4cm
{\sl Department of Physics, \\
Indian Institute of Technology,\\
Kanpur 208016, \\
India}}
\maketitle
\abstract{
\noindent
We study the phase structure and equilibrium state space geometry of R-charged black holes in $D = 5$, $4$ and $7$ and the corresponding
rotating $D3$, $M2$ and $M5$ branes. For various charge configurations of the compact black holes in the canonical ensemble we demonstrate new liquid-gas like phase 
coexistence behaviour culminating in second order critical points. The critical exponents turn out to be the same as that of four dimensional asymptotically AdS black holes in Einstein Maxwell theory. 
We further establish that the regions of stability for R-charged black holes are, in some cases, more constrained than is currently believed, due to properties of 
some of the response coefficients. The equilibrium state space scalar curvature is calculated for various charge configurations, both for the case of compact as well 
as flat horizons and its asymptotic behaviour with temperature is established.}
\end{titlepage}

\section{Introduction}

Understanding the nature of black hole thermodynamics has been the focus of intense research over the last few decades (see, eg. \cite{td1}, \cite{td2}). Whereas a 
consistent quantum theory of gravity, which is required to fully understand the microscopic basis underlying the thermodynamical description of black holes, is still lacking, 
semi-classical analyses have nevertheless provided deep insights into the rich phase structure of these systems. As is well known, these black holes exhibit phase 
transitions and critical phenomena as seen in normal thermodynamic systems. These phenomena become more important in the context of the 
AdS/CFT duality \cite{maldacena} which has, for example, led to the correspondence between the Hawking-Page phase transition in asymptotically 
AdS black holes with the confinement/deconfinement transition in the boundary field theory \cite{witten}. It is therefore important to understand the full phase structure
of AdS black holes, the study of which was initiated in \cite{gub1},\cite{caisoh},\cite{gub2},\cite{gub3}

It is well established by now that the phase structure of black holes depends crucially on the choice of the ensemble, in contrast with conventional 
thermodynamic systems. For example, it was established in \cite{johnson1}, \cite{johnson2}  that in the canonical ensemble (fixed charge) the charged Reissner-Nordstrom-AdS (RN-AdS) black holes show a first order 
liquid gas like phase transition culminating in a second order critical point, analogous to the Van der Waals gas. However, no 
such behaviour is seen in the grand canonical (fixed potential) ensemble, where, instead, a Hawking-Page phase transition occurs. In this context, we have comprehensively
analysed the Kerr-Newman-AdS (KN-AdS) black hole in \cite{tapo1} where we showed that there is a far richer phase structure for these systems than was previously known. 
Namely, we established a liquid gas like phase coexistence behaviour culminating in a second order critical point for two new ``mixed'' ensembles, wherein one 
thermodynamic charge (the electric charge or the angular momentum) and one conjugate potential (the angular velocity or the electric potential respectively) were held fixed. 
In \cite{tapo2}, we also calculated the critical exponents corresponding 
to the critical points in these ensembles, and showed that these are in fact identical to the exponents in the RN-AdS and the Kerr-AdS black holes,
suggesting an universality in the scaling behaviour of asymptotically AdS black holes in four dimensions. (See also \cite{banerjee} for additional discussions on the phase
transitions in RN-AdS black holes).

In \cite{tapo1} and \cite{tapo2}, we also analysed the case of four dimensional AdS black holes from the point of view of the intrinsic geometry of its equilibrium thermodynamic
state space (to be contrasted with the extrinsic geometric perspective of \cite{tisza},\cite{callen}). 
This intrinsic geometrical perspective of thermodynamics, pioneered by Weinhold \cite{weinhold} and Ruppeiner \cite{rupp},\cite{rupp1} has been an area of interest 
in conventional thermodynamics for decades. In \cite{weinhold}, a Riemannian metric was attributed to the equilibrium state space of a thermodynamic system. This was in terms of
the Hessian matrix of the internal energy of the system expressed as a function of the extensive variables, including the entropy. This did not, however, have a clear interpretation
in terms of a distance in the equilibrium thermodynamic state space. In \cite{rupp1}, a similar metric was introduced in terms of the Hessian of the entropy density expressed as a function of the internal energy and the other extensive variables. It could then be established that the probability distribution of thermodynamic fluctuations between equilibrium states
are related to the invariant distance between them in the thermodynamic state space. For black holes, thermodynamic geometry was first alluded to by Ferrara et al \cite{ferrara} in the 
context of extremal black holes in string theory 
and its relation to the underlying moduli space. Since then, a lot of work has been done on the application of thermodynamic geometry to black hole thermodynamics, starting from the work 
of \cite{aman}. In this context, in \cite{tapo1}, we showed that the scalar curvature of the equilibrium state space  geometry in fact captured the first order liquid-gas like phase behaviour 
of conventional Van der Waals systems, contrary to what was thought before. A detailed analysis reveals that in regions of phase coexistence, the scalar curvature shows multi-valued branch 
structures, and implies that the curvature changes branch close to the first order phase transition point. A similar behaviour of the state space scalar curvature was observed
for KN-AdS black holes in the canonical and the mixed ensembles alluded to in the previous paragraph. 
In \cite{tapo2}, the scaling behaviour of various thermodynamic quantities, including the Ruppeiner curvature was undertaken. It was established that the scalar curvature
followed a scaling law similar to the well known hyperscaling relation \cite{hes},\cite{fish}. 

One of the aims of the present work is to extend the above formalism to the case of black holes arising in gauged supergravity theories. In particular, we will be concerned
with R-charged black holes in $D=5$, $4$ and $7$, and the corresponding rotating near extremal $D3$, $M2$ and $M5$ branes. We analyse the thermodynamic
stability issues related to these black holes, and also their equilibrium state space geometry. Our results show a much richer phase structure for these systems than has
been reported, and the importance of this lies in the description of the corresponding boundary
gauge theory. Since the topic is by now well studied, let us, at the outset, summarise our main results. Firstly, dealing with black holes with compact horizon, we find that for single 
R-charged black holes in $D=5$, there is a region of phase coexistence and first order phase transitions, similar to liquid-gas
systems, that culminate at a second order critical point. We analyse this in details, and calculate the critical exponents. They turn out to be the same as those of the
four dimensional asymptotically AdS black holes discussed in \cite{tapo2}. For the case of two charged black holes in $D=5$ ( two charges are equal with the third set to zero), a similar liquid-gas like phase 
coexistence behaviour is obtained with the same set of exponents. We also analyse the three charge case (with all charges set equal). Similar phase
coexistence is established, with the same set of exponents as before.

In the grand canonical ensemble, our analysis reveals that for the single charged case discussed above for $D=5$, $4$ and $7$, the region of thermodynamic stability of the black hole 
is more constrained than what one obtains by simply considering the Hessian of the entropy function. Namely, calculating the isothermal compressibility, we find that the zeroes of these 
further constrain the thermodynamically stable region of the black hole. Interestingly, there is also a region in parameter space where the black hole exhibits negative isobaric expansivity. 
These constraints however, are not seen in the two and three charged examples. We further elucidate the equilibrium state space geometry of these black holes
for several charge configurations. The associated state space scalar curvature shows expected divergences at points of thermodynamic instability as obtained from the heat capacity corresponding to the grand canonical ensemble. 

This paper is organised as follows. In section 2, we study R-charged black holes in $D=5$. First, we deal with the compact horizon black holes with single R-charge, two charge and three charge configurations. The analyses is done both in the canonical and the grand canonical ensembles, and 
we establish the structure of the internal state space geometry for the latter. Next, we study the flat horizon case, for the same charge configurations and elucidate the nature of the 
equilibrium state space geometries for these cases. Section 3 
deals with four dimensional R-charged black holes having four R-charges. 
In section 4, we discuss our last example, namely the two R-charged black hole in seven dimensions. Section 5 concludes with discussions of our results.

\section{R-charged black holes in $AdS_5$}

A spinning D3-brane configuration is characterized by rotations in planes orthogonal to the brane forming the rotation group $SO(6)$. The three independent commuting Cartan generators of the rotation group, called the spins of the D3-brane, represent the three charges under the global $SO(6)$ $R$-symmetry group of the $\mathcal{N}=4$ field theory living on the world volume of the D3-branes. The superconformal symmetry group of the gauge theory is fully represented by the enhanced symmetries in the near horizon geometry of the near extremal D3-branes, which is described by $AdS_5 \times S^5$ supergravity, with the dual gauge theory  residing on the boundary of $AdS_5$. The Kaluza Klein reduction of the spinning D3-brane on $S^5$ results in an $\mathcal{N}=8$ $D=5$ gauged supergravity with an $SO(6)$ non-Abelian gauge group. The three independent spins in the world volume of the D3-brane therefore reduce to three $U(1)$ gauge charges of the charged black holes in the $AdS_5$ supergravity which couple to three chemical potentials. 

The metric for the R-charged black holes in $D=5$, $\mathcal{N}=8$  gauged supergravity is given by \cite{bcs},

\begin{equation}
\label{d3Rchargemetric}
ds^2=-(H_1H_2H_3)^{-2/3}f\,dt^2+(H_1H_2H_3)^{1/3}(f^{-1}\,dr^2+r^2d\Omega_{3,k})
\end{equation}

where
\begin{equation}
\label{metrlabels}
f=k-\frac{\mu}{r^2}+\frac{r^2}{l^2}\Pi_{i=1}^3\,H_i~~~;~~~H_i=1+\frac{a_i}{r^2}\,\,,\,\,i=1..3\,.
\end{equation}

Here $\mu$ is the mass parameter while $a_i$ are the charge parameters entering the metric. $k$ denotes the normalized curvature at the horizon whose position $r_+$ is obtained as the largest positive root of the equation $f=0$. For $k=1$ the line element $d\Omega_{3,k}$ is that of $S^3$ while for $k=0$ the line element is that of ${\bf R}^3$. Thus, for $k=1$ the R-charged black holes have a compact spherical horizon with the dual gauge theory living on 
${\bf R} \times S^3$, while in the case of $k=0$ the horizon is planar and infinite in extent and the corresponding gauge theory lives on ${\bf R}^4$.
We will first discuss the case $k=1$. Following \cite{gub2} we set the AdS length scale $l$ to 1 and the Newton's constant $G_5$ to $\pi/4$. The ADM mass is given as

\begin{equation}
\label{R7m}
M=\frac{3}{2}\mu+ a_1+a_2+a_3
\end{equation}
The entropy is given as
\begin{equation}
\label{R7s}
S=\frac{A}{4G_5}=2\pi\sqrt{{\Pi}_{i=1}^3\,(r_+^2+a_i)}
\end{equation}
and the charges are given as
\begin{eqnarray}
\label{R7q}
q_i^2&=&a_i(r_+^2+a_i)\left[1+\frac{1}{r_+^2}{\Pi}_{j\neq i}\,(r_+^2+a_i)\right]\nonumber\\
&=&\sqrt{a_i(\mu+a_i)}
\end{eqnarray}

The temperature and the conjugate potentials can be obtained from the above equations by using the first law of thermodynamics. The temperature is given by
\begin{equation}
\label{R7t}
T=\frac{1}{r_+^2\,A}\left[ 2r_+^6+r_+^4(1+\Sigma_{i=1}^3\,a_i)-\Pi_{i=1}^3\,a_i \right]
\end{equation}
while the chemical potentials conjugate to the charges can similarly be obtained as
\begin{equation}
\label{R7phi}
\phi_i=\frac{q_i}{r_+^2+a_i}
\end{equation}

The numerator in the expression for temperature defines the extremality condition at which the parameter $\mu$ becomes
\begin{equation}
 \mu_{crit}=2\frac{a_1a_2a_3}{r_+^2}+a_1a_2+a_2a_3+a_3a_1-r_+^4
\end{equation}
The black hole is non extremal for $\mu>\mu_{crit}$ while $\mu=0$ defines supersymmetric BPS states, \cite{gubsereff,robmyers}.\footnote{see \cite{bcs} for detailed conditions for existence of horizon for different limits of the charge parameters.} These supersymmetric states are, however, not black hole solutions since they fall in the naked singularity region. The mass corresponding to the BPS state is given by
\begin{equation}
M_{BPS}=q_1+q_2+q_3
\end{equation}
since the charges $q_i$ become equal to their respective parameters $a_i$ on setting $\mu=0$. From eq.(\ref{R7t}) it can be seen that extremal black holes exist only when all the three charges are non zero.

In the canonical ensemble all the charges $q_1,q_2,q_3$ are constrained so that the mass $M$ is the only extensive quantity the black hole exchanges with the surroundings held at a fixed temperature. The Helmholtz free energy is given as
\begin{equation}
\label{R7hehl}
F(T,q_i)=M-TS=\frac{1}{2r_+^2}\left[ r_+^4(1+\Sigma\,a_i) +r_+^2(3\Sigma_{i<j}\,a_ia_j+ 2\Sigma\, a_i)+5\Pi\,a_i-r^6  \right]
\end{equation}

In ref(\cite{gub2}), the zero of the Helmholtz free energy is regarded as that of thermal AdS with fixed R-charges. However,  thermal AdS with fixed charge cannot solve the Einstein equations and so cannot directly serve as the background action in the Euclidean path integral action calculation. In such a case one could argue that a hot gas of R-charged particles will have zero Helmholtz free energy so that the black hole is globally stable only when its free energy is negative. The local stability conditions can be suitably obtained by finding the heat capacity at constant charges, $C_Q$.

In the grand canonical ensemble, on the other hand, the black hole exchanges its $U(1)$ charges with the surrounding medium at fixed values of the gauge potentials and the temperature. The Gibbs free energy is given as
\begin{equation}
\label{R7gibb}
G=-\frac{1}{2\,r_+^2}\left[r_+^6+r_+^4(\Sigma\,a_i-1)+r_+^2(\Sigma_{i<j}\,a_ia_j)+ \Pi\,a_i \right]
\end{equation}
Since thermal AdS with a fixed pure gauge potential solves the Einstein's equations, the Euclidean action in the grand canonical ensemble can be obtained by background subtraction
of  the pure gauge thermal AdS action. As a result, the zero of the Gibbs free energy corresponds to the Hawking-Page transition between the black hole and the thermal AdS. At the same time the local stability conditions are determined by the heat capacity at constant potential $C_{\phi}$ and other susceptibilities.

In our subsequent discussion on the phase structure and stability of these R-charged black holes we shall find it convenient to consider special cases where some charges are set to zero or are set equal to each other.
The cases we will consider are
\begin{eqnarray}
\label{R7cases}
\mbox{Case 1}&&~~~a_1= a,\,a_2=a_3=0\nonumber\\
\mbox{Case 2}&&~~~a_1=a_2=a, a_3=0\nonumber\\
\mbox{Case 3}&&~~~a_1=a_2=a_3=a
\end{eqnarray}
These choices are made only to simplify the algebra, and our methods can be equally well applied to arbitrary charge configurations. 

\subsection{k=1, Case 1}

The mass ,charge and entropy can be obtained by using the condition for case 1 in eq. (\ref{R7cases}) into eqs. (\ref{R7m}),  (\ref{R7q}) and (\ref{R7s}),
\begin{equation}
\label{R7case1m}
M=\frac{3}{2}\,{r_+}^{2}+\frac{3}{2}\,{r_+}^{4}+\frac{3}{2}\,a{r_+}^{2}+a
\end{equation}

\begin{equation}
\label{R7case1q}
q=\sqrt {a \left( {r_+}^{2}+a \right)  \left( 1+{r_+}^{2} \right) }
\end{equation}

\begin{equation}
\label{R7case1s}
S=2\,\pi \,{r_+}^{2}\sqrt {{r_+}^{2}+a}
\end{equation}

It can be verified that for all positive values of $r_+$ and $a$, the black hole remains within the BPS bound while  the bound gets saturated for $r_+=0$ which corresponds to a naked singularity.

The temperature and potential may be obtained similarly or by using first law of thermodynamics.

\begin{equation}
\label{R7case1t}
T=\frac{1}{2\pi}\,{\frac {1+2\,{r_+}^{2}+a}{\sqrt {{r_+}^{2}+a} }}
\end{equation}

\begin{equation}
\label{R7case1phi}
\phi={\frac {\sqrt {a}\sqrt {1+{r}^{2}}}{\sqrt {{r}^{2}+a}}}
\end{equation}

Let us consider the canonical ensemble first. The Helmholtz free energy for the first case is reduced to the following
\begin{equation}
\label{R7case1f}
F=\frac{1}{2}\,{r_+}^{2}-\frac{1}{2}\,{r_+}^{4}+\frac{1}{2}\,a{r_+}^{2}+a
\end{equation}
while the heat capacity at constant charge $C_q$ is obtained as
\begin{equation}
\label{R7case1Cq}
C_q=2\pi\,{\frac { \left( 1+2\,{r_+}^{2}+a \right)  \left( 3\,a{r_+}^{2}+4\,a+3\,{r_+}^{2}+3\,{r_+}^{4} \right) \sqrt {{r_+}^{2}+a} }{6\,a+5\,a{r_+
}^{2}+{r_+}^{2}+2\,{r_+}^{4}-{a}^{2}-1}}
\end{equation}

For the canonical ensemble it will be useful to invert eq.(\ref{R7case1q}) and obtain the parameter $a$ in terms of the charge $q$
\begin{equation}
\label{R7case1a}
a=\frac{1}{2}\,{\frac {\sqrt {{r_+}^{4}+2\,{r_+}^{6}+{r_+}^{8}+4\,{q}^{2}+4\,{r_+}^{2}{q}^{2}}-{r_+}^{2}-{r_+}^{4}}{1+{r_+}^{2}}}
\end{equation}
Using eq.(\ref{R7case1a}) we can directly express the thermodynamic variables obtained above in terms of $q$ and $r_+$ which will help us in obtaining phase plots.

The local stability condition for the canonical ensemble in terms of the charge $q$ and the horizon radius $r_+$ is obtained from the expression for the heat capacity $C_q$ in eq.(\ref{R7case1Cq}), by substituting the value of $a$ from (\ref{R7case1a}). Thus the heat capacity $C_q$ is positive for the range of values of charge given as 
\begin{eqnarray}
\label{R7case1local}
q\leq\sqrt {17\,{r_+}^{2}+17+3\,\sqrt {33\,{r_+}^{4}+64\,{r_+}^{2}+32}} \left( 1+{r_+}^{2} \right)~~~\rm{and}\nonumber\\
q\geq\sqrt {17\,{r_+}^{2}+17-3\,\sqrt {33\,{r_+}^{4}+64\,{r_+}^{2}+32}} \left( 1+{r_+}^{2} \right)
\end{eqnarray}

$C_q$ diverges along curves where the inequalities are saturated.

The zeros of the Helmholtz free energy may be similarly obtained in terms of $q$ and $r_+$. The black hole is globally stable for the range of charge
\begin{equation}
\label{R7case1hfree}
q\leq{\frac {{r_+}^{2}\,\sqrt {2\,{r_+}^{6}+{r_+}^{4}-2\,{r_+}^{2}-1}}{{r_+}^{2}+2}}
\end{equation}

\begin{figure}[t!]
\begin{minipage}[b]{0.5\linewidth}
\centering
\includegraphics[width=3in,height=2.5in]{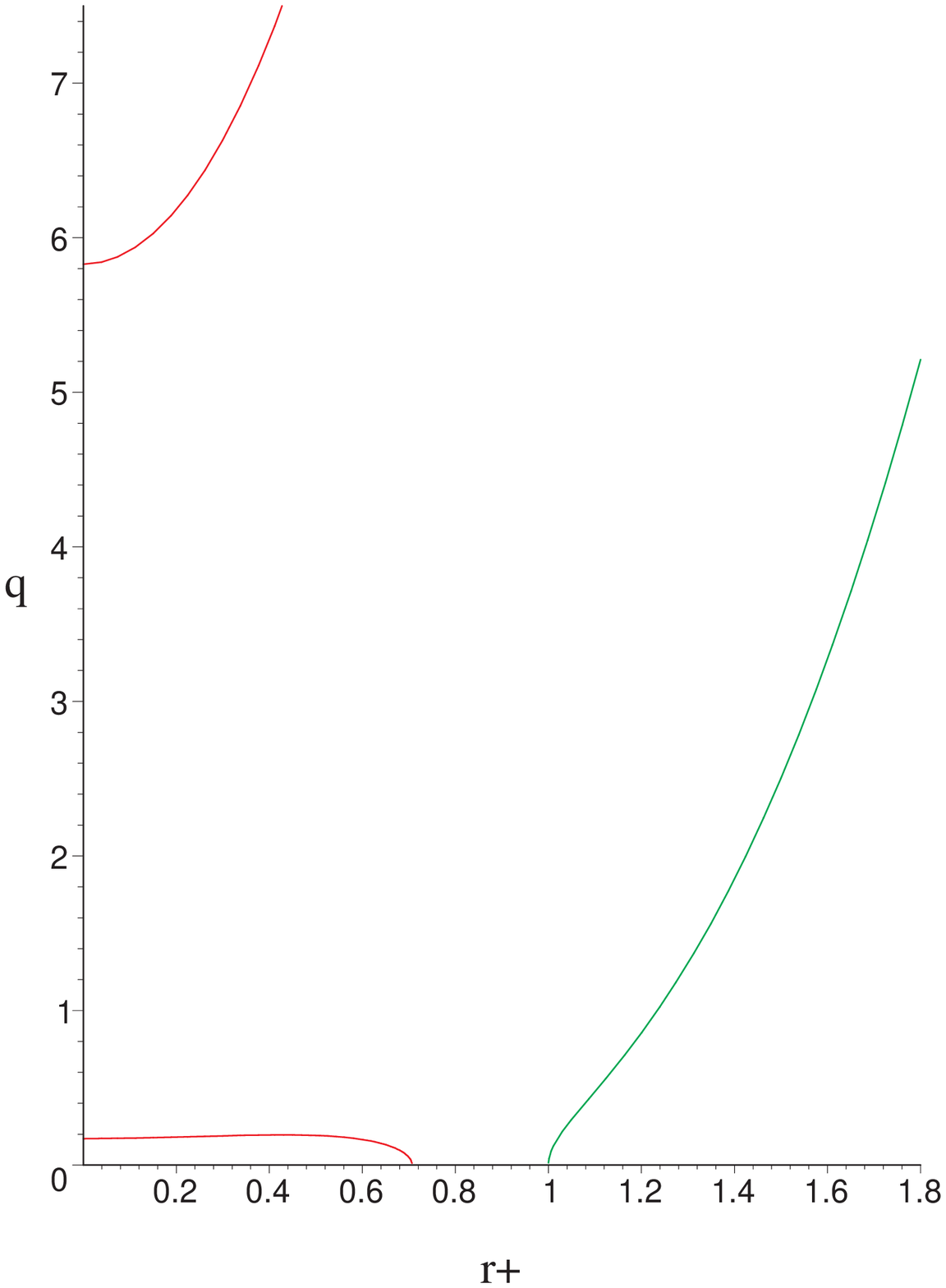}
\caption{Canonical phase diagram for the single charge D=5 R-charge black hole in $q$-$r_+$ plane. The red colored local stability curves are the infinities of $C_q$ while the green colored global stability curve is the zero of the free energy $F$. }
\label{R7c1f1}
\end{minipage}
\hspace{0.6cm}
\begin{minipage}[b]{0.5\linewidth}
\centering
\includegraphics[width=2.7in,height=2.7in]{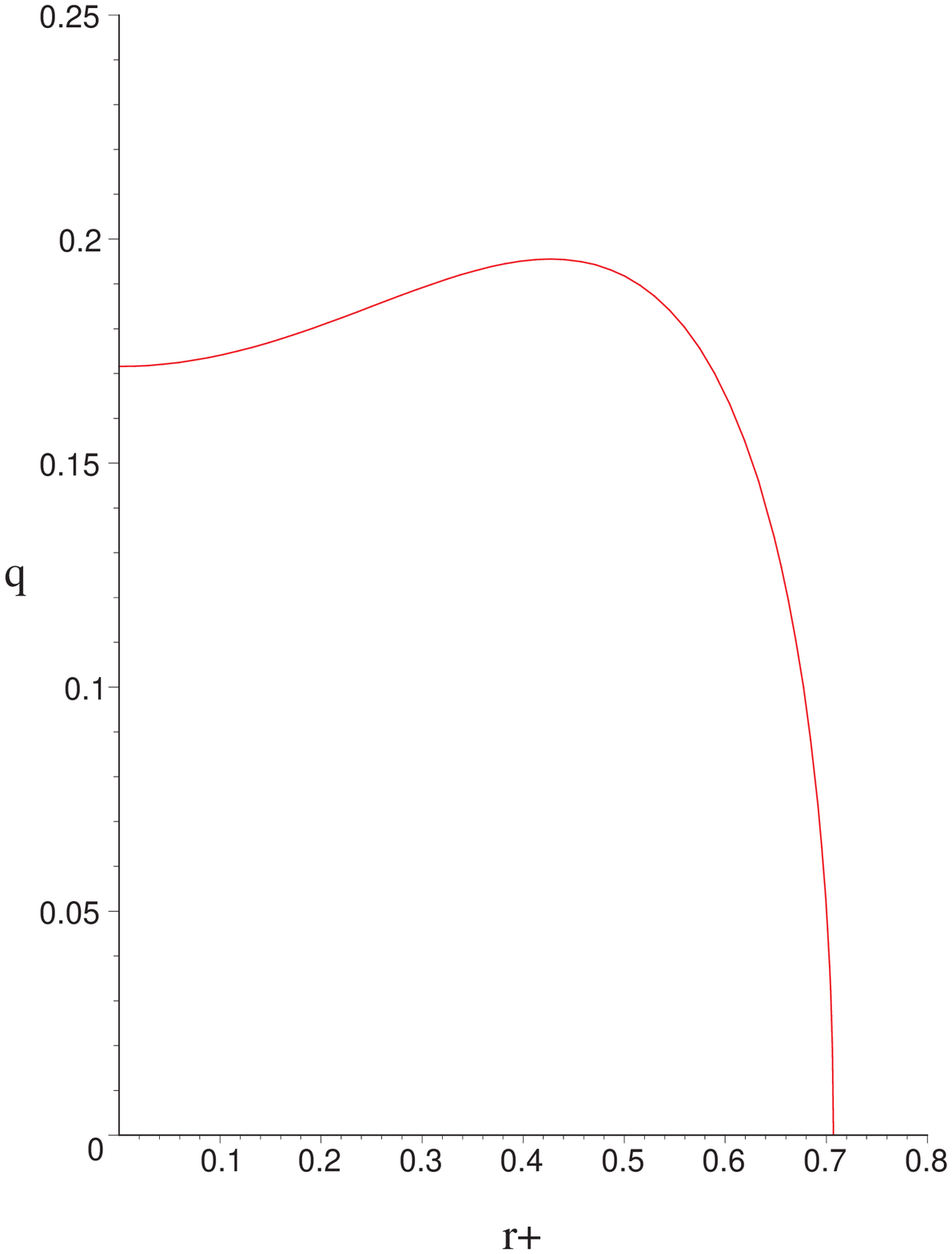}
\caption{Close up of the lower branch of the red colored local stability curve of fig.(\ref{R7c1f1}). For $0.171<q<0.195(q_c)$ a locally stable small black hole branch can exist for certain temperatures. $q=q_c$, $r=0.428$ is the critical point.}
\label{R7c1f2}
\end{minipage}
\end{figure}

Using eq.(\ref{R7case1local}) and eq.(\ref{R7case1hfree}) the phase structure for the canonical ensemble can be obtained in the $q$-$r_+$ plane. In fig.(\ref{R7c1f1}) we reproduce the canonical phase diagram of \cite{gub2} in the $q$-$r_+$ plane.
The heat capacity $C_q$ is positive in the region between the two red colored local stability curves, while the Helmholtz free energy is negative in the region below the green colored curve. The upper branch of the local stability curve meets the $y$-axis at $q_1=5.83$. The lower branch of the local stability curve has been drawn separately in fig.(\ref{R7c1f2}). It meets the $y$-axis at $q_2=0.172$, and has a maxima at $q_3=0.1955$. This indicates an interesting liquid-gas like phase coexistence behaviour in the system. 
Let us see if we can substantiate this. 

\begin{figure}[!t]
\centering
\includegraphics[width=3in,height=2.5in]{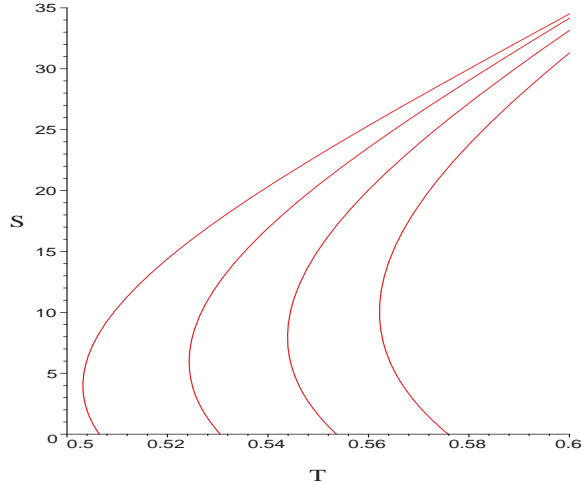}
\caption{$S$ vs $T$ plot the canonical ensemble of the single charge black hole with $q > q_1 = 5.83$. We have taken $q=8,\,9,\,10,\,11$. Similar curves can be obtained for the 
case $q < q_2 = 0.172$}
\label{R7c1st1}
\end{figure}

The canonical ensemble phase behaviour can be classified into four distinct regimes based on whether (\rmnum{1}) $q>q_1$, (\rmnum{2}) $q_1>q>q_3$, (\rmnum{3}) $q_3>q>q_2$ and (\rmnum{4}) $q<q_2$. From figs(\ref{R7c1f1}) and (\ref{R7c1f2}) it can be seen that for the first and the fourth cases the black hole is locally unstable for values of $r_+$ starting from zero and up to the point of intersection of the constant $q$ curves (\emph{i.e}, the horizontal lines) with the stability curves. On an entropy-temperature plane, as in fig.(\ref{R7c1st1}), where we have plotted isocharge $S$ vs $T$ curves for 
$q > q_1$ , the unstable small black hole branch coexists with the stable large black hole branch starting from the turning point temperature corresponding to the divergence of $C_q$ up to the temperature
\begin{equation}
T_1=\frac{(1+q)}{2\pi\sqrt {q}}
\label{R7t1}
\end{equation}

The temperature $T_1$ is obtained by setting $r_+=0$ in eq.(\ref{R7case1t}) and it corresponds to the BPS state. For $T>T_1$ only the locally stable branch exists. On further increasing the temperature, constant $q$ lines (cf. fig. (\ref{R7c1f1})) cross the free energy curve and the black hole becomes becomes globally stable. 

\begin{figure}[t!]
\begin{minipage}[b]{0.5\linewidth}
\centering
\includegraphics[width=3in,height=2.5in]{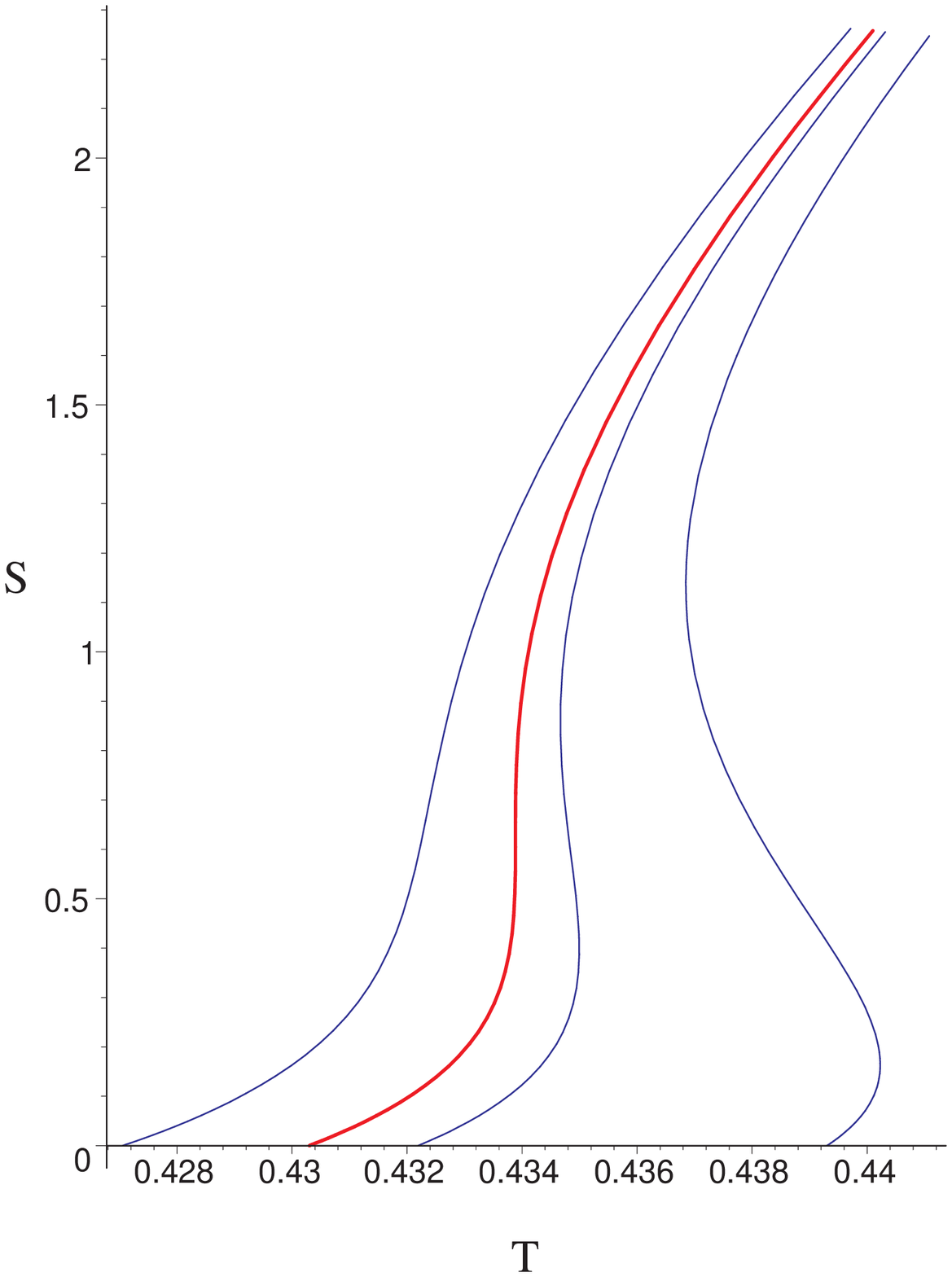}
\caption{$S$ vs $T$ plots of constant charge curves in the canonical ensemble of the single charge case with $q$ near $q_c$. From right to left the constant charge curves have $q=0.184,\,0.193,\,0.1955(q_c),\,0.2.$ }
\label{R7c1st2}
\end{minipage}
\hspace{0.6cm}
\begin{minipage}[b]{0.5\linewidth}
\centering
\includegraphics[width=2.7in,height=2.7in]{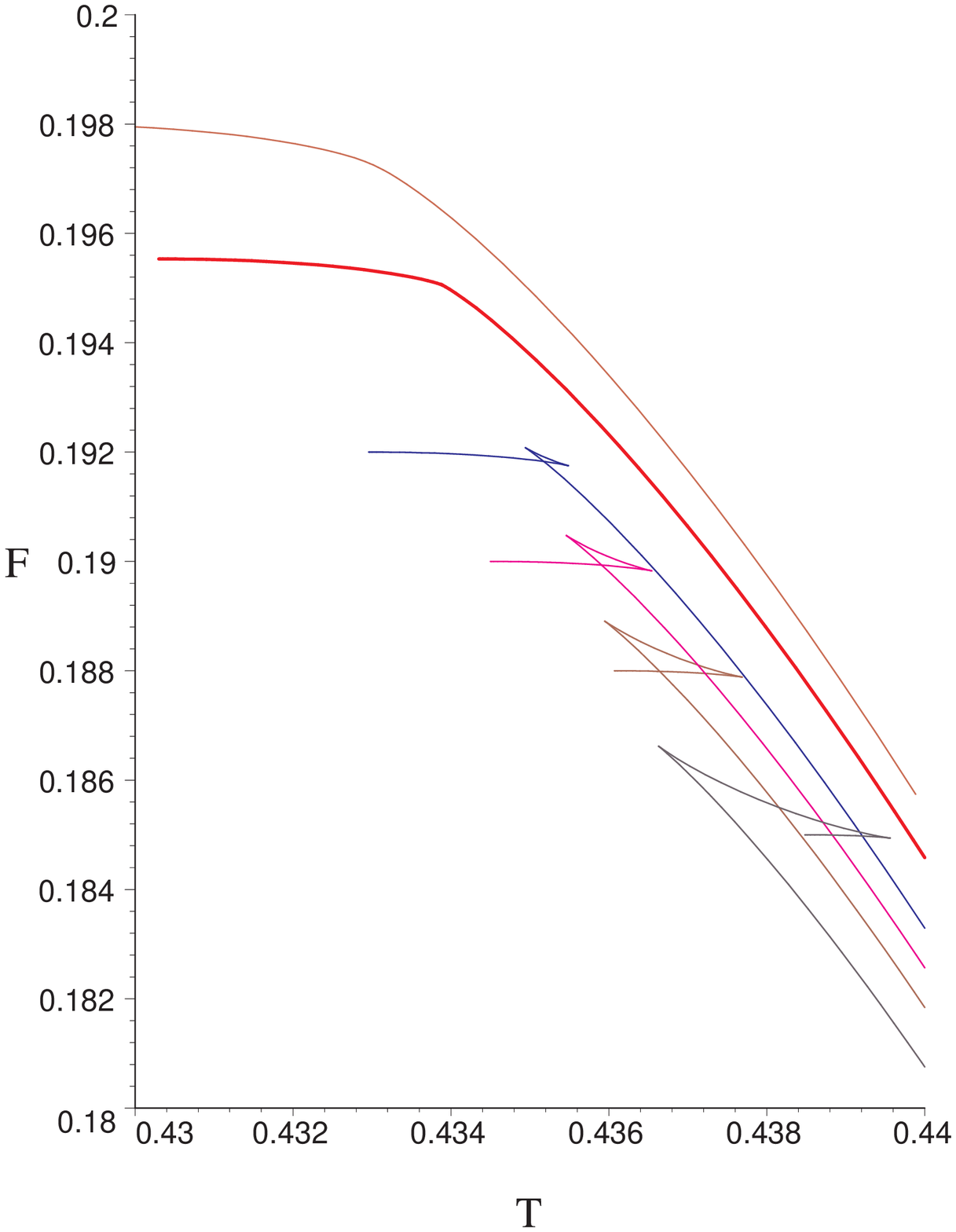}
\caption{$F$ vs $T$ plots of constant charge curves with $q$ near $q_c$. From the bottom to top the free energy curves are at charges $q=0.184$, $0.188$, $0.190$, $0.192$, $0.1955(q_c)$, and, $0.198$ respectively.}
\label{R7c1ft1}
\end{minipage}
\end{figure}

For case (\rmnum{2}) there is only one stable branch for all temperatures starting with a naked singularity at $T=T_1$. The third case is the most interesting since in this range of charge there is a stable small black hole (sbh) branch separated from the stable large black hole (lbh) branch by an unstable branch as can be seen from fig.(\ref{R7c1f2}). This is
because $C_q$ is negative within the lower red stability curve. 
Moreover, the constant charge line $q=q_3$ becomes tangent to the stability curve in fig.(\ref{R7c1f2}). This means that the heat capacity $C_q$ remains positive all along the $q_3$ line and diverges at the tangent point, clearly indicating a second order phase transition between the sbh and the lbh branches.
In fig.(\ref{R7c1st2}) we plot the constant charge curves in the $S$-$T$ plane with charge in the vicinity of $q_3$. The critical curve (in thick red) has a point of inflection at $T=T_c=0.434$. For $q>q_3$ there is a single stable black hole branch corresponding to case (\rmnum{2}). We further investigate this phase coexistence behaviour by drawing the Helmholtz free energy with temperature in fig.(\ref{R7c1ft1}). The swallow tail shape of the free energy curve indicating phase coexistence behaviour is apparent. We further observe an ``incomplete'' swallowtail at the bottom of figure. This shows that for small enough charges the free energy of the sbh branch remains higher than that of the lbh branch so that for such charges there will be no first order transition between the two branches. The swallow tail becomes ``complete'' at $q_4\sim 0.1866$ as we have checked numerically. Thus, for $0.1955>q>0.1866$ the black holes show first order transition between the sbh and the lbh branches culminating in a critical point. From fig.(\ref{R7c1ft1}) it can also be seen that the critical temperature lies to the left of the first order transition temperatures on the temperature axis.

We now briefly discuss the scaling behaviour near the critical point in the canonical ensemble. For the canonical ensemble the order parameter can be chosen to be the radius $r_+ $ or, equivalently, following \cite{johnson1}, the potential $\phi$. Note that in the canonical ensemble the appropriate isothermal susceptibility corresponding to the critical exponent $\gamma$ will be $\kappa'_T=(\partial\phi/\partial q)_T$ instead of its inverse $\kappa_T=(\partial q/\partial \phi)_T$ \cite{wu}. The critical exponents $\alpha $ and $\beta $ are obtained by first going to the constant charge lines in the $q-r_+ $ plane of fig.(\ref{R7c1f2}). This is facilitated by converting the equations for $T$, eq.(\ref{R7case1t}), and $C_q$, eq.(\ref{R7case1Cq}), in terms of $q$ and $r_+ $ by using eq.(\ref{R7case1a}) as already mentioned. By Taylor expanding in powers of $r_+ $ on the line $q=q_c$ around the critical point $(q_c,{r_+}_c)=(0.196,0.428)$ we may check that to the leading order
\begin{equation}
\label{R7crit1}
T-T_c\sim r_+^3~~~;~~~C_q^{-1}\sim r_+^2
\end{equation}

In order to obtain the critical exponents $\gamma$ and $\delta$ for the susceptibility $\kappa'_T$ and the order parameter $\phi$ we invert the equation for the temperature eq.(\ref{R7case1t}) and obtain the parameter $a$ in terms of $T$ and $r_+$. Two branches are obtained
\begin{eqnarray}
a_1=-2\,{r_+}^{2}-1+2\,{\it T}\, \left( {\it T}\,\pi +\sqrt {{{\it T}}^{2}{\pi }^{2}-1-{r_+}^{2}} \right) \pi\nonumber\\
a_2=-2\,{r_+}^{2}-1+2\,{\it T}\, \left( {\it T}\,\pi -\sqrt {{{\it T}}^{2}{\pi }^{2}-1-{r_+}^{2}} \right)
\end{eqnarray}

Of these the second branch $a_2$ is the one relevant to the critical point. Expressing $\kappa'_T$, $\phi$ and $q$ in terms of $T$ and $r_+$ we Taylor expand these in powers of $r_+$ around the critical point $(T_c,{r_+}_c)=(0.434,0.428)$. It turns out that to the leading order
\begin{equation}
\label{R7crit2}
q-q_c\sim r_+^3~~~;~~~\phi-\phi_c\sim r_+~~~;~~~{\kappa'_T}^{-1}\sim r_+^2
\end{equation}

Using eq.(\ref{R7crit1}) and eq.(\ref{R7crit2}) we obtain the critical exponents as

\begin{equation}
\label{R7cancr}
{\alpha}=2/3, ~~{\beta}=1/3,~~ {\gamma}= 2/3,~~ {\delta}=3 ~.
\end{equation}

These turn out to be exactly the same as those for the canonical ensemble of the RN-AdS \cite{johnson2},\cite{wu} and the Kerr-AdS black holes or the two mixed ensembles
of the KN-AdS black holes obtained in \cite{tapo2}.

We now investigate the phase structure of single charge black holes in the grand canonical ensemble. For this ensemble the Gibbs energy reduces to
\begin{equation}
G=-\frac{1}{2}\,{r_+}^{2} \left( -1+r_+^{2}+a \right)
\end{equation}

Since the grand canonical ensemble is characterized by $T$ and $\phi$ as the independent control parameters it will be convenient, as an intermediate step, to invert eq.(\ref{R7case1phi}) and express the parameter $a$ in terms of $\phi$,
\begin{equation}
\label{R7cs1aphi}
a={\frac {{r_+}^{2}{\phi}^{2}}{-{\phi}^{2}+1+{r_+}^{2}}}
\end{equation}
This equation also implies that for non-negative $a$,
\begin{equation}
r_+^2\geq \phi^2-1
\end{equation}
which means that while for $\phi\leq1$ the horizon radius has a minimum at $r_+=0$ at which $a$ is zero, for $\phi>1$ the horizon radius has a finite minimum at which $a$ is infinite.
Using eq.(\ref{R7cs1aphi}), the temperature may be expressed in terms of $\phi$ and $r_+$ as

\begin{equation}
\label{R7cs1tphi}
T=\frac{1}{2\pi}\,{\frac {\sqrt {1+{r_+}^{2}} \left( 2\,{r_+}^{2}-{\Phi}^{2}+1 \right) }{r_+\sqrt {-{\Phi}^{2}+1+{r_+}^{2}} }}
\end{equation}

\begin{figure}[t!]
\begin{minipage}[b]{0.5\linewidth}
\centering
\includegraphics[width=3in,height=2.5in]{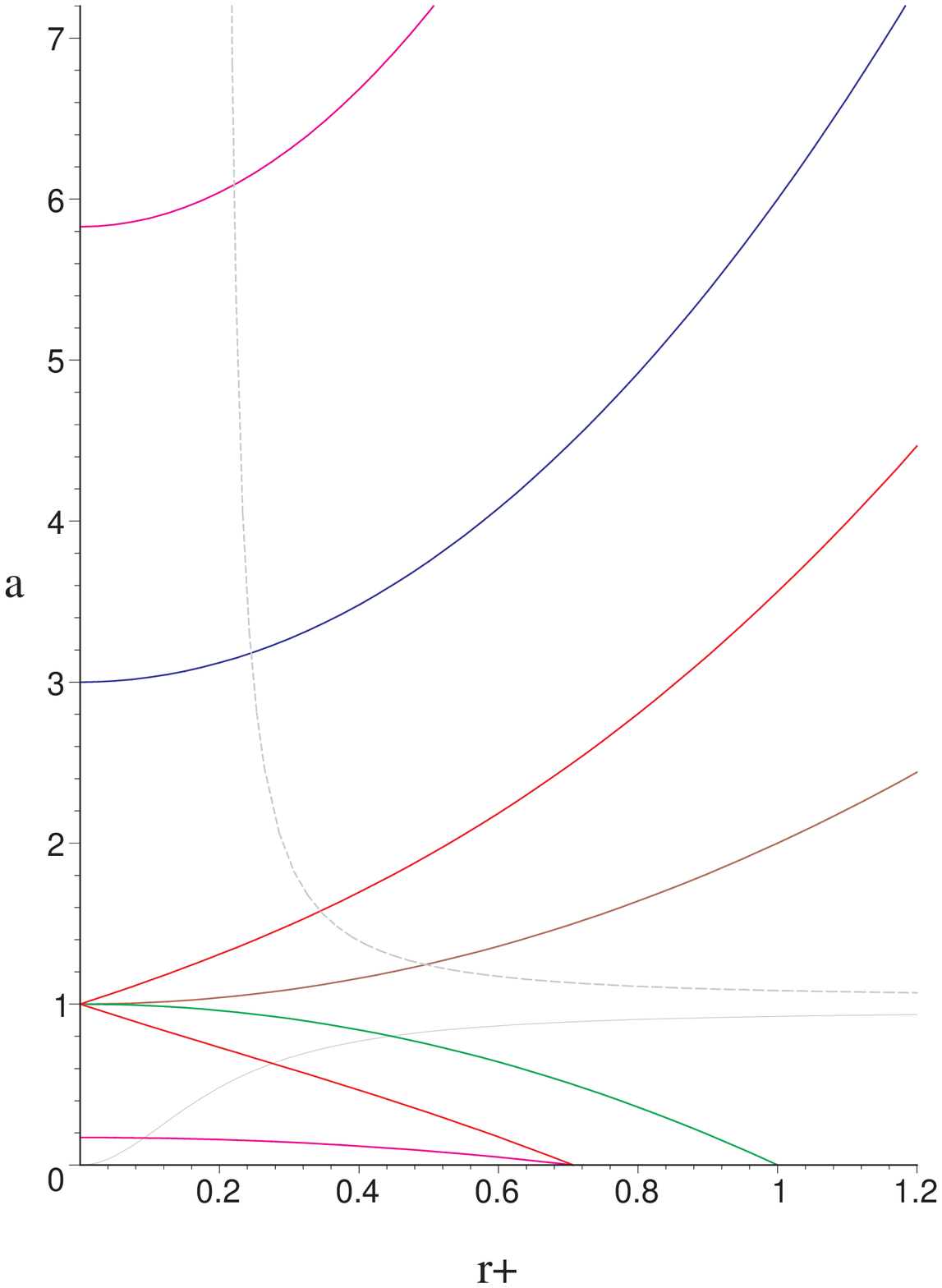}
\caption{Stability curves of $C_\phi, \kappa_T,\alpha_\phi$ in the $a-r_+$ plane in the grand canonical ensemble of the single charge case. The curve of zeros of Gibbs energy is colored green. The grey curves are isopotentials, with the lower one at $\phi=0.98$ and the upper one at $\phi=1.02$ }
\label{R7c1ph1}
\end{minipage}
\hspace{0.6cm}
\begin{minipage}[b]{0.5\linewidth}
\centering
\includegraphics[width=2.7in,height=2.7in]{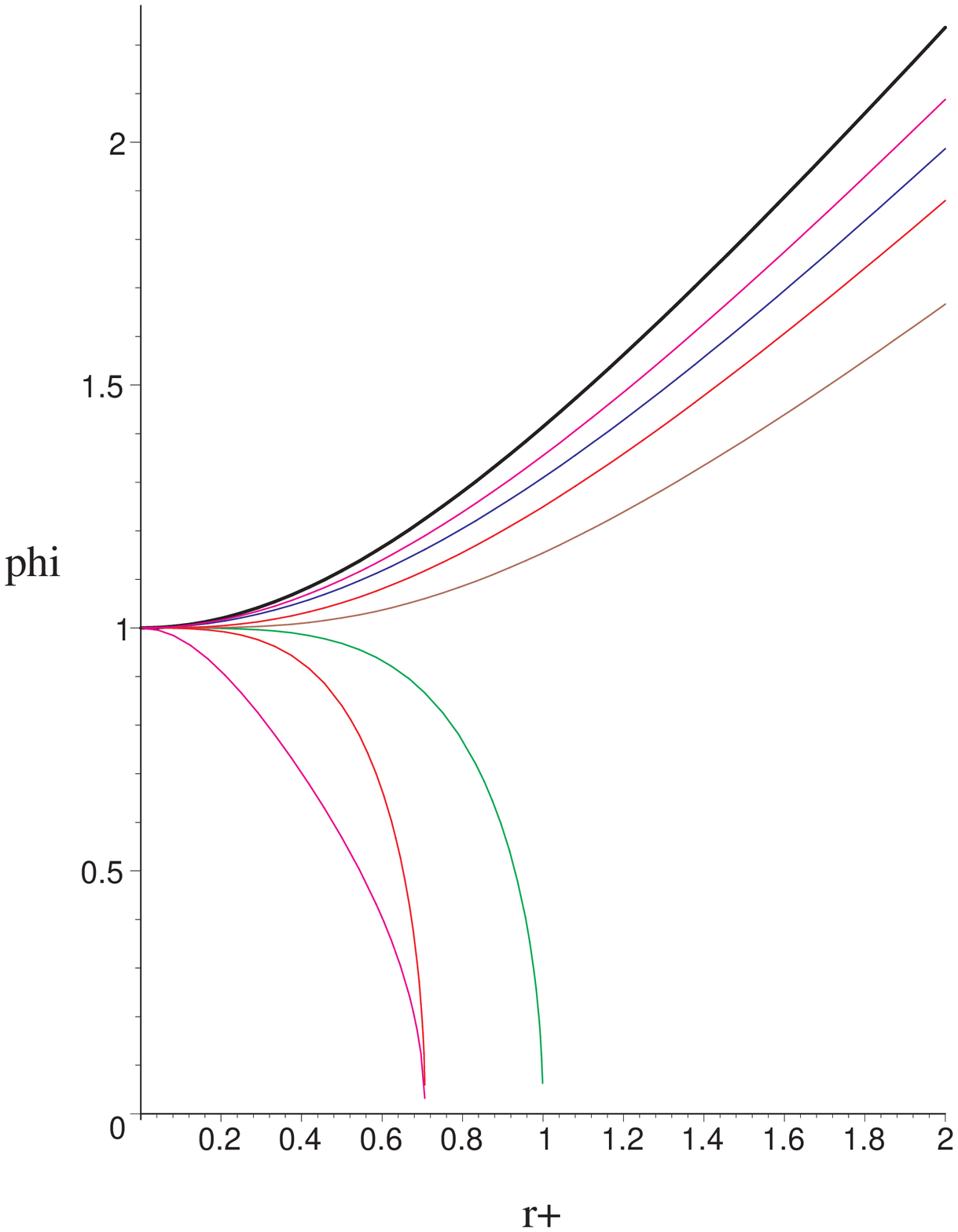}
\caption{$r_+-\phi$ plane plot of infinities of $C_\phi$ in red, zeros of $C_\phi$ in blue, zeros of Gibbs energy in green. The black curve corresponds to the physical limit $r_+=\sqrt{\phi^2-1}$ on which $a$ is infinite and above which $a$ becomes negative.}
\label{R7c1ph2}
\end{minipage}
\end{figure}

The local stability condition can be determined by the positivity of the heat capacity $C_\phi$ and the susceptibilities, $(\partial q/\partial\phi)_T$ and  $(\partial q/\partial T)_\phi$. The heat capacity $C_\phi$ is obtained as
\begin{equation}
\label{R7case1cphi}
C_\phi= 2\,{\frac {\sqrt {{r_+}^{2}+a} \left( 1+2\,{r_+}^{2}+a \right) \pi \,{r_+}^{2} \left( a-3\,{r_+}^{2}-3 \right) }{{a}^{2}+1-2\,a-{r_+}^{2}a-{r_+}^{2}-2
\,{r_+}^{4}}}
\end{equation}

and the susceptibilities are obtained as follows
\begin{equation}
\label{R7case1kappa}
\kappa_T=\left(\frac{\partial q}{\partial\phi}\right)_T={\frac { \left( {r_+}^{2}+a \right)  \left( {a}^{2}-6\,a-5\,{r_+}^{2}a-{r_+}^{2}-2\,{r_+}^{4}+1 \right) }{-2\,a-{r_+}^{2}a+{a}^{2}+1-{r_+}^{2}-2\,{r_+}^{4
}}}
\end{equation}

\begin{equation}
\alpha_\phi=\left(\frac{\partial q}{\partial\,T}\right)_\phi=4\,{\frac { \left( {r_+}^{2}+a \right) \sqrt {a}\sqrt {1+{r_+}^{2}} \left( -a+1+{r_+}^{2} \right) \pi }{2\,a+{r_+}^{2}a-{a}^{2}-1+{r_+}^{2}+2\,
{r_+}^{4}}}
\end{equation}

Let us note that while the negative sign of the isothermal capacitance $\kappa_T$ does indicate an electrical instability in the black hole the same for $\alpha_\phi$ does not point to any instability as such. Indeed, charge is conjugate to the potential and not temperature.
  
The zeros and the infinities of $C_\phi,\kappa_T,\alpha_\phi$ and the zeros of $G$ have been plotted together in the $a-r_+$ plane of fig.(\ref{R7c1ph1}). The two magenta colored curves corresponding to the zeroes of $\kappa_T$ are seen to be the same as the stability curves corresponding to the divergence of $C_q$ in fig.(\ref{R7c1f1}). The two red curves correspond to the infinities of $C_\phi$, $\kappa_T$ and $\alpha_\phi$ since their divergence is governed by the same polynomial. The blue curve corresponds to the zeros of $C_\phi$ while the brown curve corresponds to the zeros of $\alpha_\phi$. Finally, the green curve corresponds to the zeros of the Gibbs free energy. 
$\kappa_T$ is positive below the lower magenta curve, above the upper
magenta curve, and between the two red curves. $C_\phi$ is positive between
the two red curves and above the blue curve. $\alpha_\phi$ is positive
between the brown and the lower red curve and above the upper red curve. However, as mentioned above, the sign of $\alpha_\phi$ is not an indicator of a thermodynamic instability. It only indicates the different responses of the black hole in the two regions.
It can therefore be deduced that that the black hole achieves complete
local thermal and electrical stability in two regions. One of them lies
above the upper magenta curve while the other one lies in between the
upper red curve and the lower red curve.
Global stability is attained in regions outside the green curve. Further, as may be verified from eq.(\ref{R7cs1aphi}), $\phi=1$ corresponds to the line $a=1$ in the $a-r_+$ plane, so that it demarcates the $\phi<1$ regions below from the $\phi>1$ regions above. Moreover, the $\phi=1$ line is fully thermodynamically stable (both locally and globally) for all temperatures starting with a BPS solution at $r_+=0$ and $T=1/\pi$, \cite{gub2}. It can be seen that all the constant $\phi$ curves asymptote to the $a=1$ line for large $r_+$. Two grey colored in curves have also been plotted in the figure, with the lower one at potential $\phi=0.98$ and the upper one at $\phi=1.02$.

In fig.(\ref{R7c1ph2}) we plot the stability curves in the $\phi-r_+$ plane, with the stability curves retaining the color coding of fig.(\ref{R7c1ph1}). The boundary of the physical region is indicated by the black curve $\phi=\sqrt{r_+^2+1}$ on which the parameter $a$ is infinite, and above which it becomes negative. 
As before, the zeros of the Gibbs free energy are shown in green and correspond to the Hawking-Page transition between the black hole and thermal AdS. The regions of complete local thermodynamical stability lie between the upper red curve and the lower red curve on the one hand and between the black curve and the upper magenta curve on the other hand. Since $G$ is positive below and negative above the green curve it can be clearly seen that in the region bounded by the lower red curve and the green curve the black hole is locally thermodynamically stable but globally unstable, in other words, it is metastable. Also notice that for $\phi>1$ the black hole is always globally stable.
It seems therefore that the condition for thermodynamical stability
requires more information than can be obtained from the negativity condition for the Hessian of entropy.
This is because the Hessian provides only the red curves (infinities of
$C_\phi$)from its determinant and the magenta curves from its principal
minor (infinities of $C_q$) as shown in fig.(\ref{R7c1ph1}) and
fig.(\ref{R7c1ph2}). Evidently, the zeros of $C_\phi$, shown by the blue
curve in the two figures, does not appear as a condition for the negativity of the
Hessian, even though it plays an important role in determining the phase structure.

\begin{figure}[t!]
\begin{minipage}[b]{0.5\linewidth}
\centering
\includegraphics[width=3in,height=2.5in]{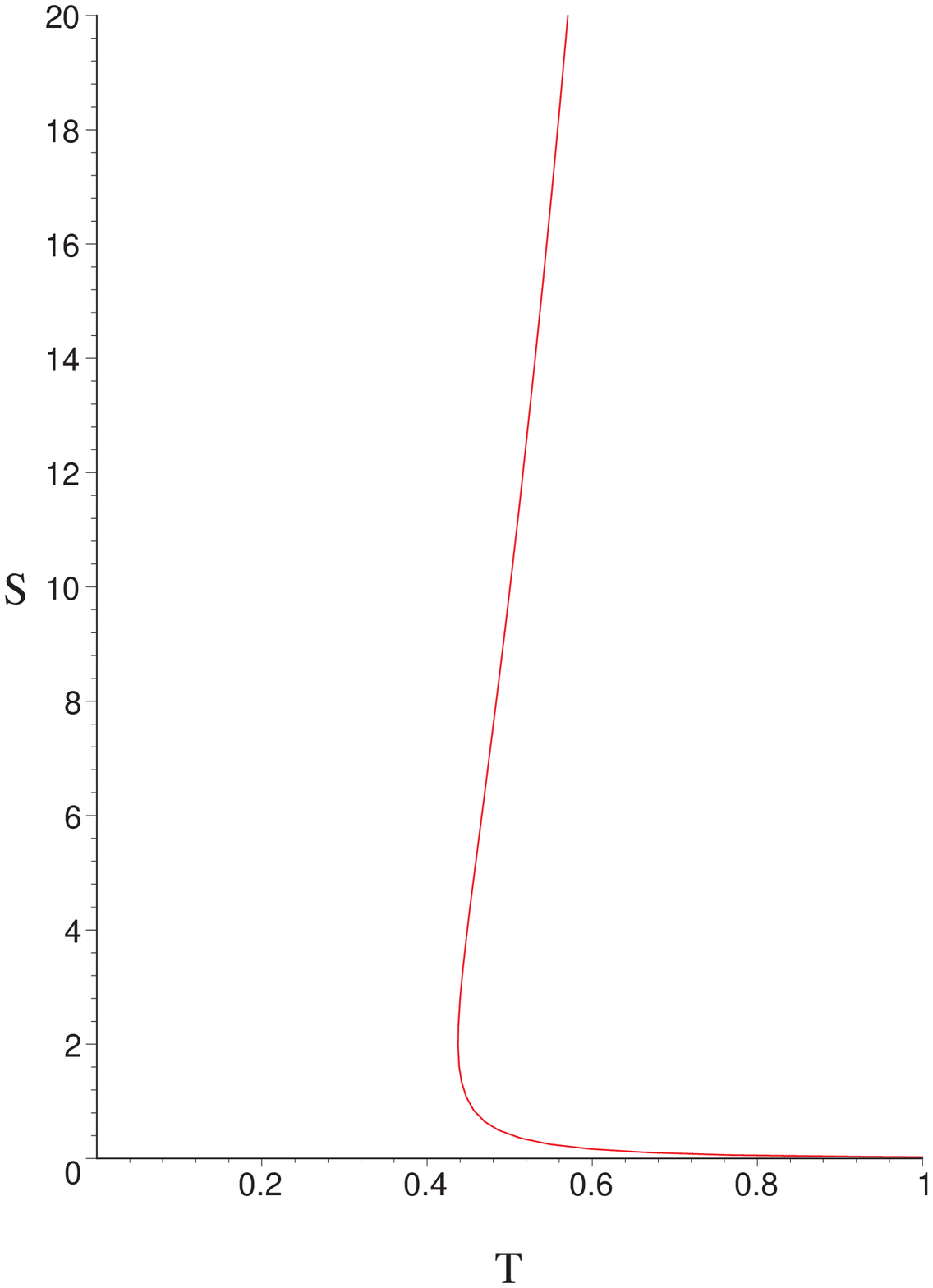}
\caption{The figure shows an isopotential plot of entropy $S$ with the temperature $T$ for the single R-charged black hole in $D=5$, with the potential fixed at $\phi=0.4$. At the turning point $T=0.44$ and $r_+=0.67$.}
\label{R7w}
\end{minipage}
\hspace{0.6cm}
\begin{minipage}[b]{0.5\linewidth}
\centering
\includegraphics[width=2.7in,height=2.7in]{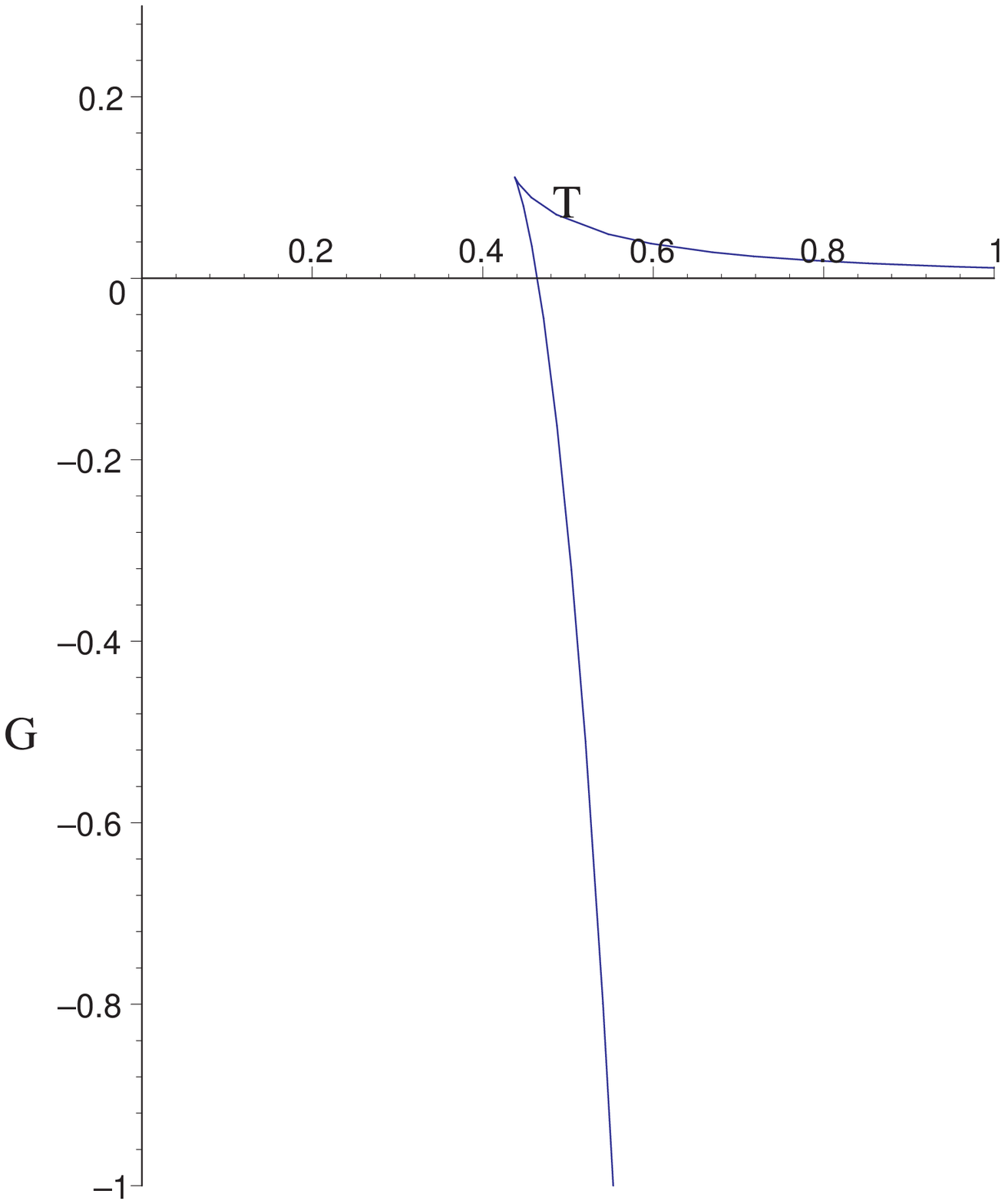}
\caption{The figure shows an isopotential plot of the Gibbs energy $G$ vs $T$, with the potential $\phi=0.4$. The locally stable branch is the one with the lower free energy and it crosses zero at $T_{HP}=0.46$.}
\label{R7x}
\end{minipage}
\end{figure}

\begin{figure}[t!]
\begin{minipage}[b]{0.5\linewidth}
\centering
\includegraphics[width=3in,height=2.5in]{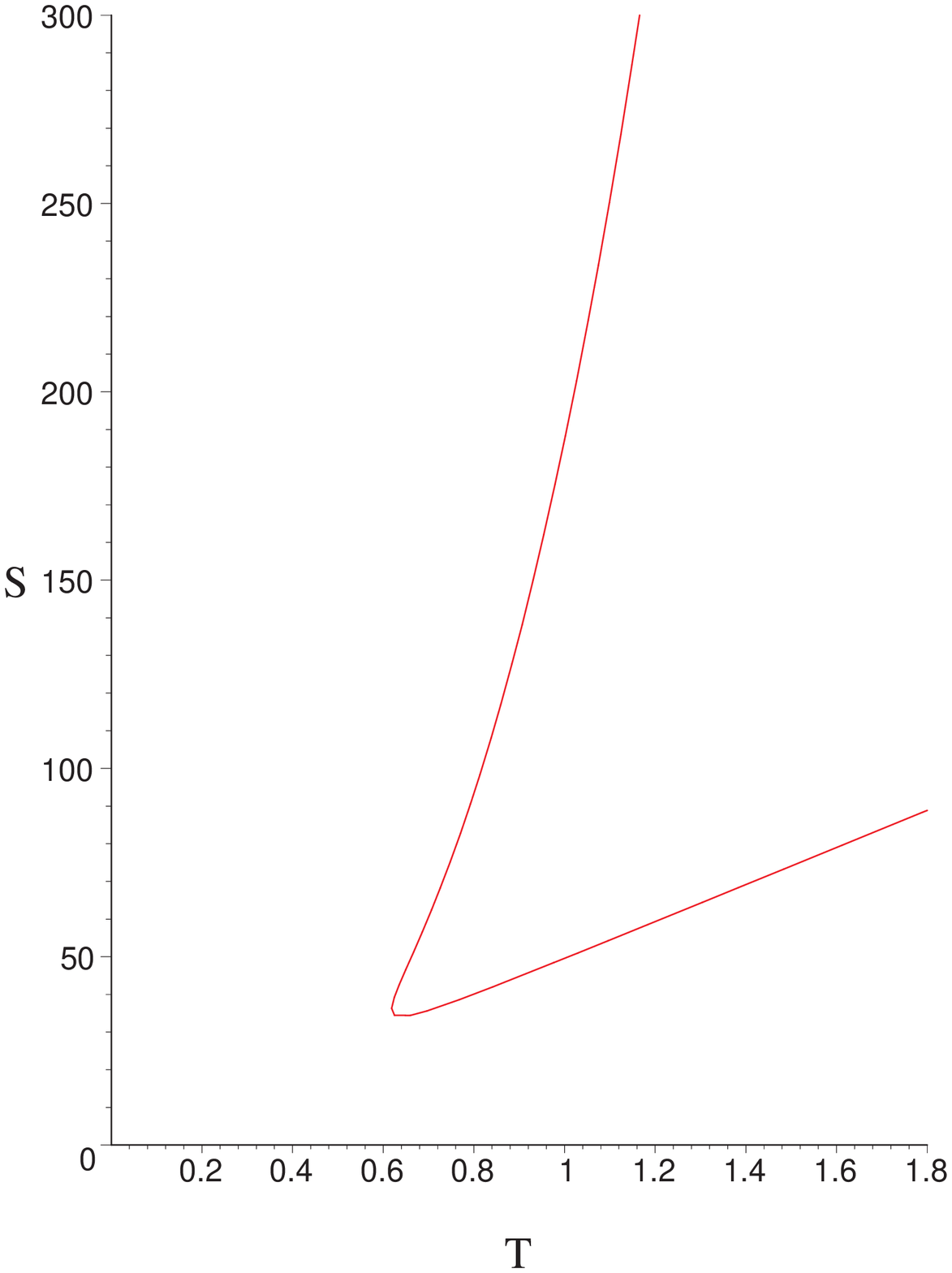}
\caption{Isopotential plot in $S-T$ plane at $\phi=1.5$. At the turning point (the Davies point), $T_D=0.62$ and $r_+=1.43$. Complete thermodynamic stability is obtained 
on the lower branch for $T\geq T_2=0.72$.}
\label{R7y}
\end{minipage}
\hspace{0.6cm}
\begin{minipage}[b]{0.5\linewidth}
\centering
\includegraphics[width=2.7in,height=2.7in]{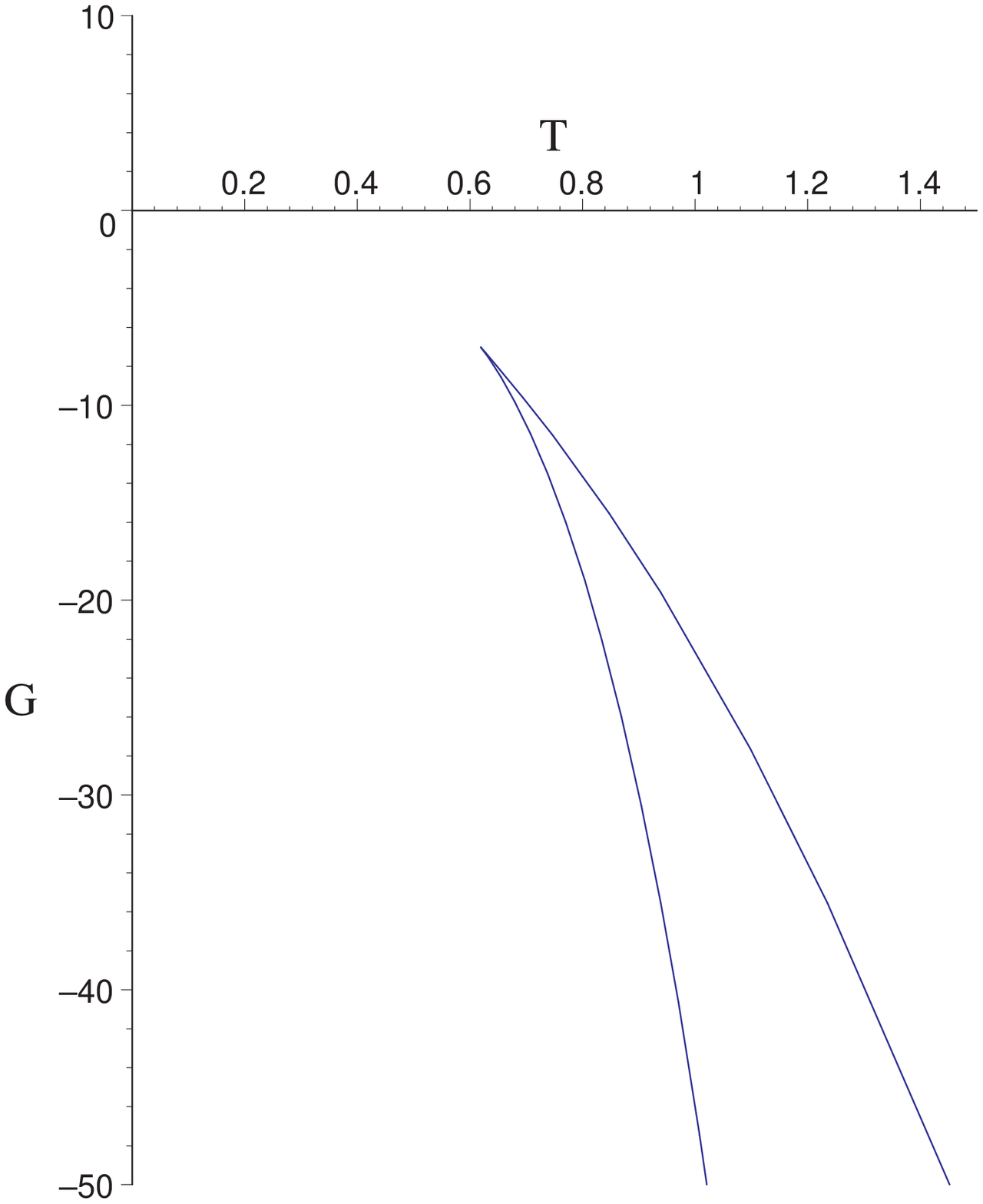}
\caption{$G$ vs $T$ plot at $\phi=1.5$. The lower free energy branch corresponds to the upper locally thermally stable branch of fig.(\ref{R7y}).}
\label{R7z}
\end{minipage}
\end{figure}

We further illustrate the phase behaviour in the grand canonical ensemble by obtaining constant potential curves. Fig.(\ref{R7w}) and fig.(\ref{R7x}) represent a typical isopotential curve in the $S-T$ and $G-T$ plane respectively for $\phi<1$. The phase behaviour is similar to the grand canonical ensemble of the RNAdS black hole as discussed in \cite{tapo2}. Fig(\ref{R7y}) shows a typical isopotential curve in $S-T$ plane with $\phi>1$. The lower branch achieves full thermodynamical stability for $T>T_2$ (c.f caption of fig. (\ref{R7y})) which corresponds to crossing to left of the upper magenta curve in fig.(\ref{R7c1ph2}). Whereas on the upper branch both the horizon radius and the charge become large with increasing temperature, on the lower branch only the charge becomes large while the radius $r_+$ decreases and approaches $r=\sqrt{\phi^2-1}$ as $T$ approaches infinity. In fig.(\ref{R7z}) we draw isopotential curves of $G$ vs. $T$ with the same potential as in fig.(\ref{R7y}). The lower branch of Gibbs free energy corresponds to the upper branch of of $S$ vs $T$ curve in fig.(\ref{R7y}). From the figure it is apparent that the black hole always has a negative Gibbs energy for $\phi>1$ so that the temperature of the black hole formation itself could be considered as the Hawking Page temperature. For $\phi=1$ there is only one branch and it is locally as well as globally stable for all values of $r_+$ as can easily be inferred from fig.(\ref{R7c1ph2}). Further, from eq.(\ref{R7cs1tphi}) we can see that the $\phi=1$ black hole solution starts at a finite temperature $T_1=1/\pi$ at $r_+=0$ which, it can be verified, is the lower limit of black hole temperature for any potential, \cite{yamada}.

We now briefly discuss the thermodynamic geometry associated with these single charged black holes. The thermodynamic line element is positive in regions where the Hessian of the entropy is negative. However, as we have seen, for the single charged black hole there are regions of state space that are thermodynamically stable in spite of them not meeting the requirement of stability according to the Hessian condition. In other words, thermodynamic stability does not imply the positivity of the line element in this case.
\begin{figure}[t!]
\centering
\includegraphics[width=3in,height=2.5in]{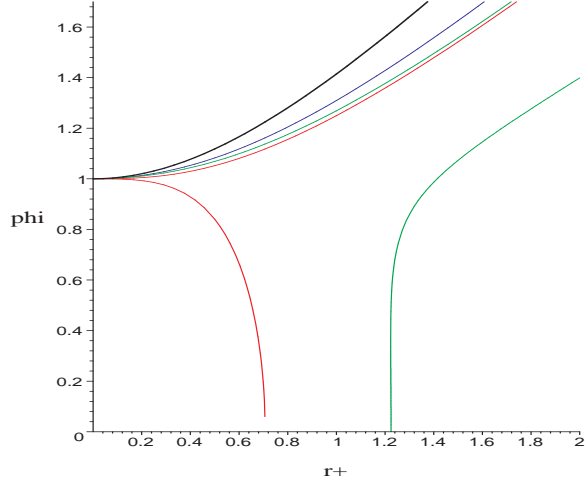}
\caption{ $\phi-r_+$ plot showing zeros (in green) and infinities (in red) of the state space curvature for a single R-charged black hole in $D=5$. The blue curve indicates the zeros of $C_\phi$ while the black curve corresponds to the physical limit in the $\phi-r_+$ plane.}
\label{fig00c}
\end{figure}

The state space scalar curvature for the single non zero charge case is obtained as follows

\begin{eqnarray}
R &=& 
\frac{\left(1 + 7r_+^2 + 6a + 6r_+^4 + 3r_+^2a -3 a^2\right)\left(-3a^2 + 2a - 3 - r_+^2a - r_+^2 +2r_+^4\right)}{\pi\left( 2a + r_+^2a - a^2 -1 + r_+^2 + 2r_+^4\right)^2}\nonumber\\
&\times& \frac{\left(1+ r_+^2\right)\sqrt{r_+^2 + a}}{\left(4a + 3r_+^2a + 3r_+^4 + 3r_+^2\right)\left(1 + 2r_+^2 + a\right)}
\label{singlechargeRcase1}
\end{eqnarray}

\begin{figure}[h!]
\begin{minipage}[b]{0.5\linewidth}
\centering
\includegraphics[width=3in,height=2.5in]{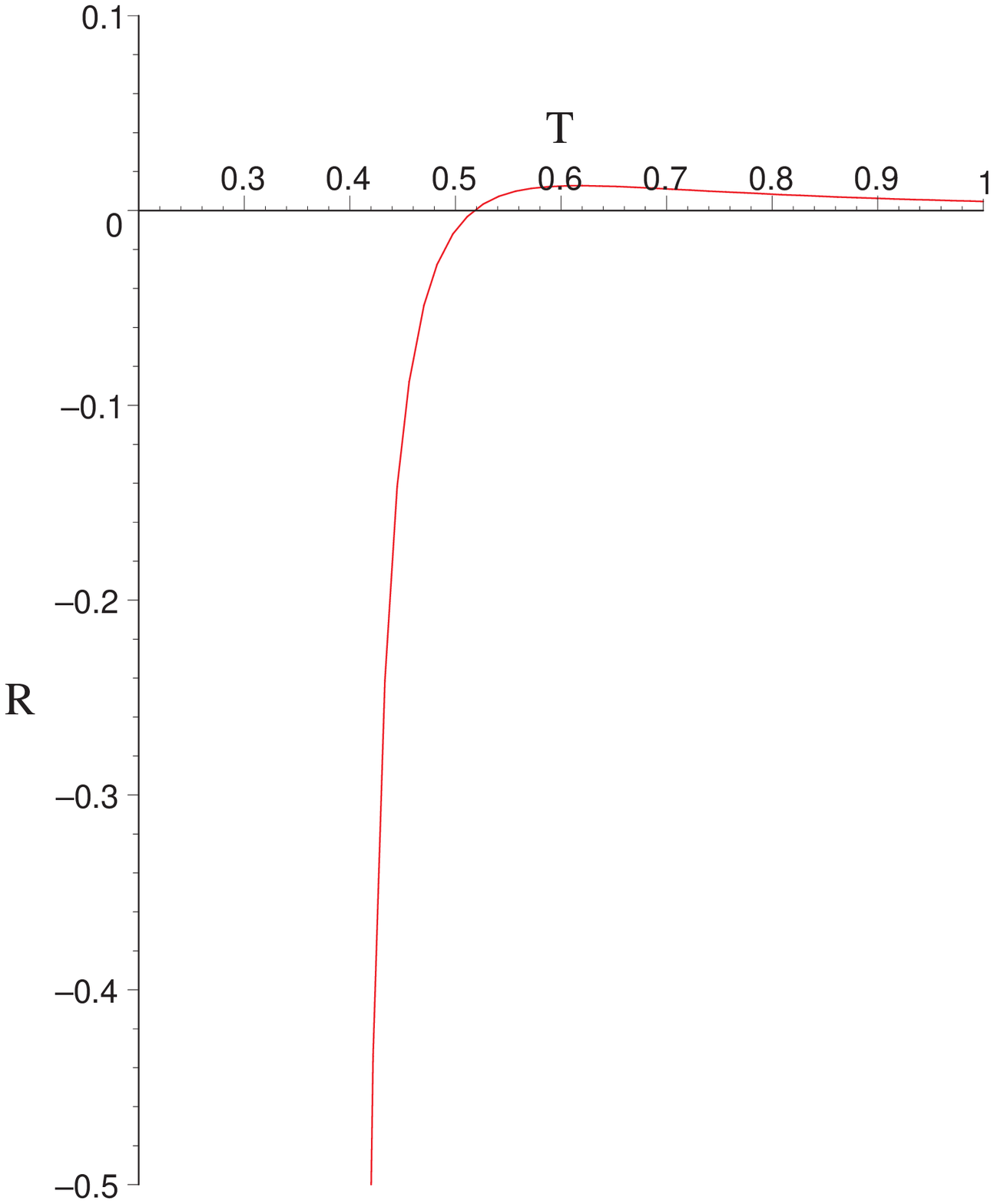}
\caption{Isopotential plot of $R$ vs $T$ for the single charge case with $\phi=0.8$ }
\label{R7c1R1a}
\end{minipage}
\hspace{0.6cm}
\begin{minipage}[b]{0.5\linewidth}
\centering
\includegraphics[width=2.7in,height=2.7in]{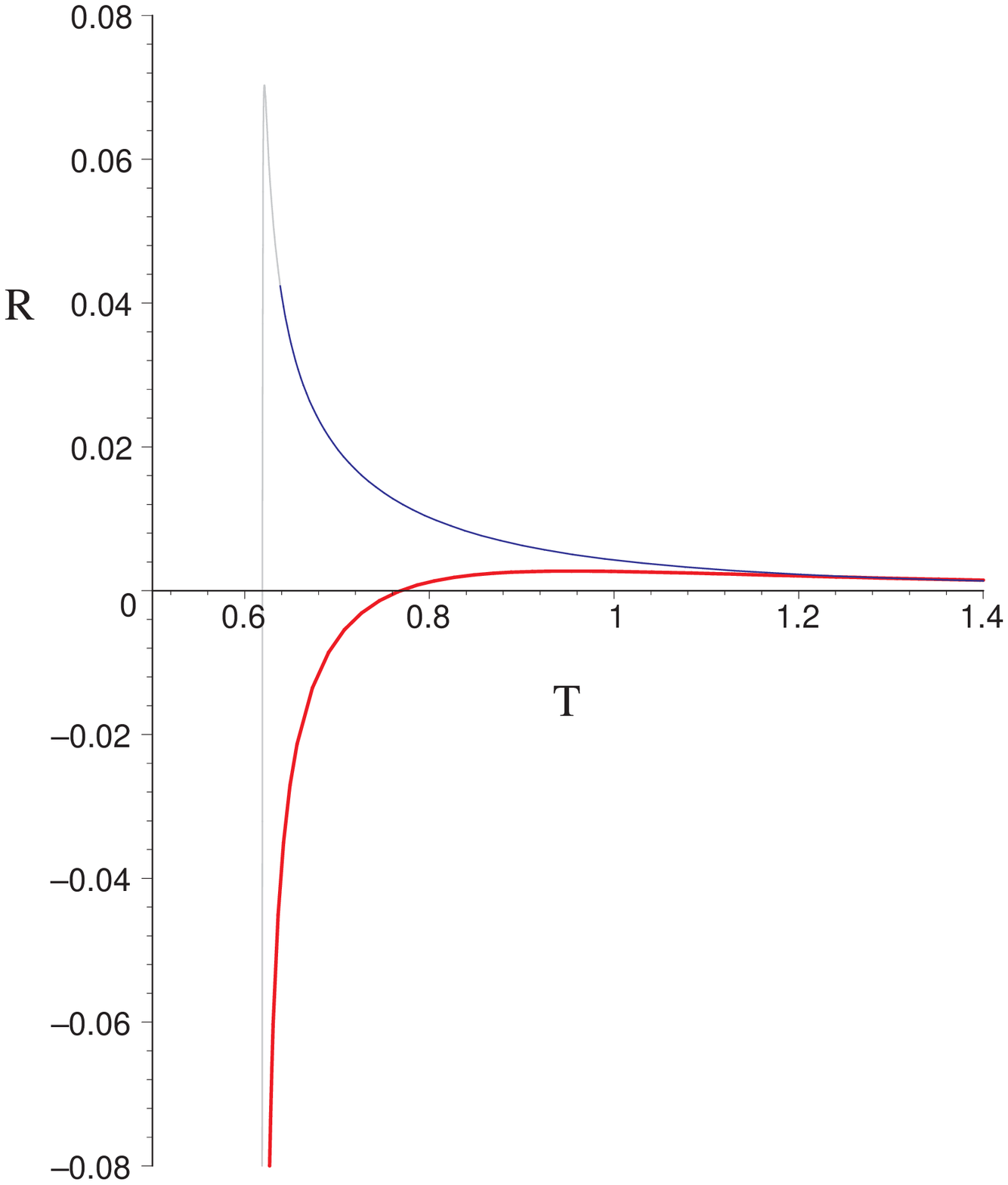}
\caption{Isopotential plot of $R$ vs $T$ for the single charge case with $\phi=1.5$.}
\label{R7c1R1b}
\end{minipage}
\end{figure}

Expectedly, the curvature diverges in the same regions as the heat capacity $C_\phi$ or the susceptibilities $\kappa_T$ and $\alpha_\phi$. In fig.(\ref{fig00c}) we plot the zeros of $R$ in green and the infinities of $R$ (which are the same as the infinities of $C_\phi$) in red. The zeros of $C_\phi$ are also included in the plot and are shown in blue. Contrary to the RN-AdS case the zeros of $R$ bear no relation to the zeros of the Gibbs free energy. Ignoring the detailed phase picture as in fig.(\ref{R7c1ph2}) we focus here only on the thermal stability. For $\phi<1$ thermodynamic geometry is not defined in regions inside the lower red curve, which corresponds to the thermally unstable small black hole branch of fig.(\ref{R7w}), since the line element is negative there. For the thermally stable large black hole branch the state space curvature starts with a negative divergence and becomes positive on crossing the green curve. A typical such curve has been shown in the  $R-T$ plane in fig.(\ref{R7c1R1a}). For the case $\phi>1$, as mentioned earlier, the region between the black curve and the blue curve corresponds to a thermally stable black hole, the region between the blue curve and the upper red curve corresponds to the thermally unstable small black hole while the region to the right of the upper red curve is the thermally stable large black hole. A typical plot of $R$ vs $T$ for $\phi>1$ is shown in fig.(\ref{R7c1R1b}) where the red colored branch corresponds to the thermally stable large black hole while the blue colored branch corresponds to the thermally stable part of the small black hole branch. 
An interesting observation from fig.(\ref{fig00c}) is that for $\phi>1$ the red curve
corresponding to the infinity of $R$ is closely aligned to one of the
green curves corresponding to the zero of $R$. In fact, it can be shown
that in the large black hole limit the two curves coincide. This leads to
a simplification in the expression for the curvature of flat horizon black
holes as will become clear in a later discussion on the black holes with
k=0.
Admittedly, there seems to be no relation between the zeros of the Gibbs free energy and the zeros of the curvature.

The asymptotic behaviour of the state space scalar curvature at high temperatures can also be determined. 
It can be verified that for both the large black hole branch and the stable small black hole branch, the decrease of the state space curvature at large temperatures is given by
\begin{equation}
\label{Rdecay1}
R\sim \frac{1}{T^3}
\end{equation}

\subsection{k=1, Case 2}

In this case two of the three $U(1)$ charges become equal while the third becomes zero so that we have $q_1=q_2=q,q_3=0$. The first law of thermodynamics for this case becomes
\begin{equation}
\label{R7cs21stlaw}
dM=TdS+2\phi dq
\end{equation}

Since in this case fluctuations in the equal charges are also the same we can as well interpret this to be a black hole system with charge $2q$. Absorbing the factor of two into the definition of charge the first law can be rephrased as
\begin{equation}
\label{R7cs21stlawa}
dM=TdS+\phi dq
\end{equation}

We list the thermodynamic quantities for these ``2-charged'' black holes. The mass becomes
\begin{equation}
\label{7cs1M}
M=\frac{3}{2}\,{r_+}^{2}+\frac{3}{2}\,{r_+}^{4}+3\,a{r_+}^{2}+\frac{3}{2}\,{a}^{2}+2\,a
\end{equation}
while the entropy and charge are obtained as
\begin{equation}
\label{7cs1S}
S=2\,\pi \, \left( {r_+}^{2}+a \right) r_+
\end{equation}
and
\begin{equation}
\label{7cs1Mq}
q=2\,\sqrt {a}\sqrt {{r_+}^{2}+a}\sqrt {1+{r_+}^{2}+a}
\end{equation}

The temperature and the electric potential are similarly given as
\begin{equation}
\label{7cs1T}
T=\frac{1}{2}\,{\frac {r_+ \left( 1+2\,{r_+}^{2}+2\,a \right) }{\pi \, \left( {r_+}^{2}+a \right) }}
\end{equation}
and
\begin{equation}
\label{7cs1phi}
\phi={\frac {\sqrt {a}\sqrt {1+{r_+}^{2}+a}}{\sqrt {{r_+}^{2}+a}}}
\end{equation}
Interestingly, in this case the temperature goes to zero at the naked singularity as can be checked by setting $r_+=0$ in eq.(\ref{7cs1S}) and eq.(\ref{7cs1T}).

Let us now discuss the phase behaviour of these two charged black holes. We consider the canonical ensemble first. The Helmholtz free energy for the two charge case is given by

\begin{equation}
\label{7cs1F}
F=\frac{1}{2}\,{r_+}^{2}-\frac{1}{2}\,{r_+}^{4}+a{r_+}^{2}+\frac{3}{2}\,{a}^{2}+2\,a
\end{equation}

The heat capacity at constant charge $C_q$ is obtained as
\begin{equation}
\label{7cs1Cq}
C_q=2 \pi \, {\frac {r_+ \left( 1+2\,{r_+}^{2}+2\,a \right)\left( {r_+}^{2}+a \right)  \left( 3\,{a}^{2}+6\,a{r_+}^{2}+2\,a+3\,{r_+}^{4}+3\,{r_+}^{2}
 \right) }{6\,{a}^{3}+14\,{a}^{2}{r_+}^{2}+7\,{a}^{2}+2\,a+8\,a{r_+}^{2}+
10\,a{r_+}^{4}+{r_+}^{4}+2\,{r_+}^{6}-{r_+}^{2}}}
\end{equation}

We can invert eq.(\ref{7cs1Mq}) for $q$ and express $a$ in terms of $q$ and $r_+$. The expression is lengthy and we will not reproduce it here. Owing to the algebraic difficulty involved, it is difficult to obtain the divergence in $C_q$, eq.(\ref{7cs1Cq}), in terms of $q$ and $r_+$ in a closed form by using the inversion just mentioned. However, instead we can obtain $a$ as a function of $r_+$ by solving for the zero of the denominator of $C_q$ which controls its divergence. Plugging the expression for $a$ thus obtained into eq.(\ref{7cs1Mq}) for $q$ we obtain  the stability curve in $q-r_+$ plane. In an analogous manner the curve for the zeros of $F$ may be obtained.

\begin{figure}[t!]
\begin{minipage}[b]{0.5\linewidth}
\centering
\includegraphics[width=3in,height=2.5in]{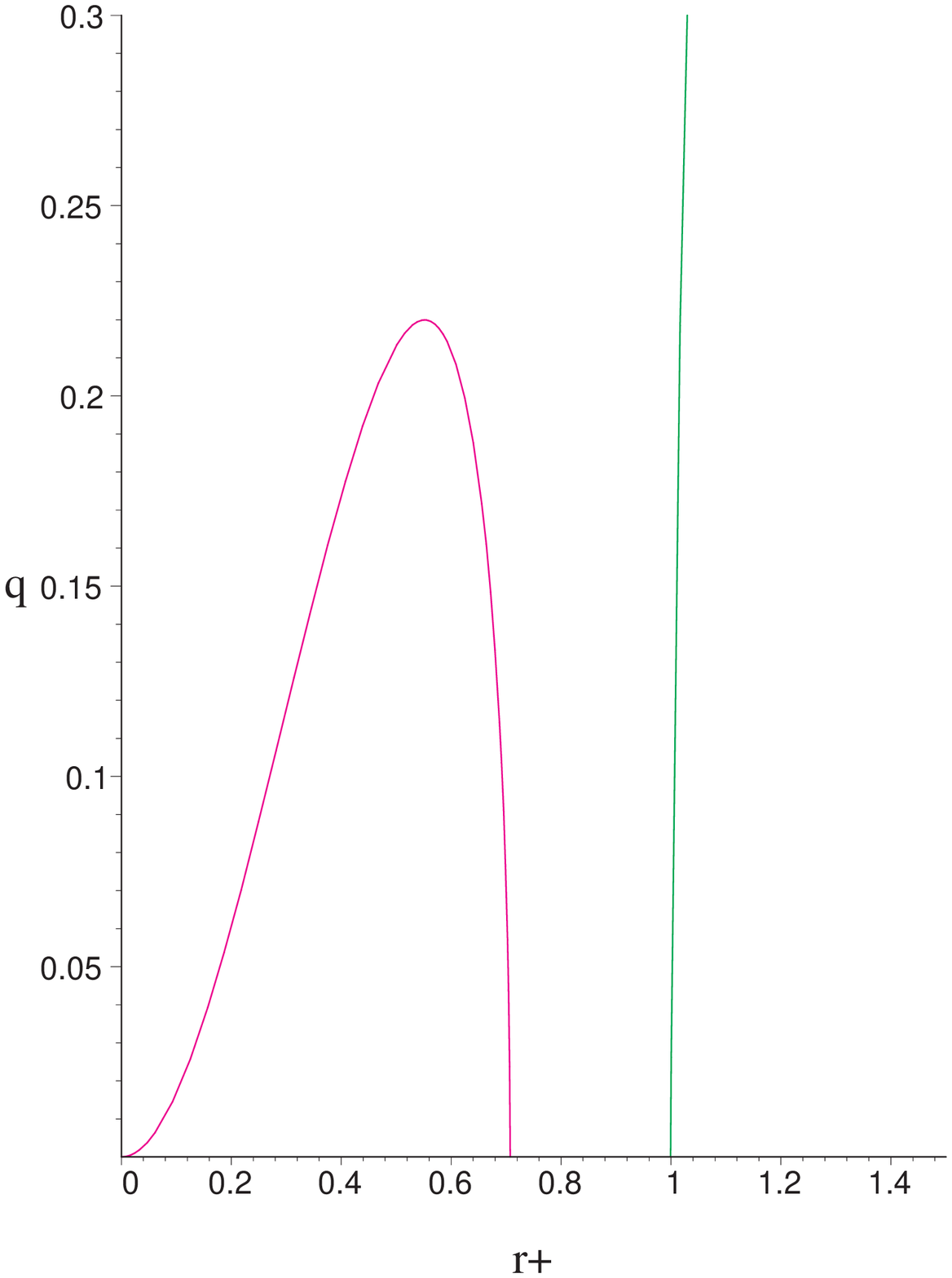}
\caption{Phase structure in the $q-r_+$ plane for the canonical ensemble in the two charge case. The stability curve showing infinities of $C_q$ is magenta colored with the maxima at $q=q_c=0.220$ while the zeros of Helmholtz energy are colored green.}
\label{R7cs2ph}
\end{minipage}
\hspace{0.6cm}
\begin{minipage}[b]{0.5\linewidth}
\centering
\includegraphics[width=3in,height=2.5in]{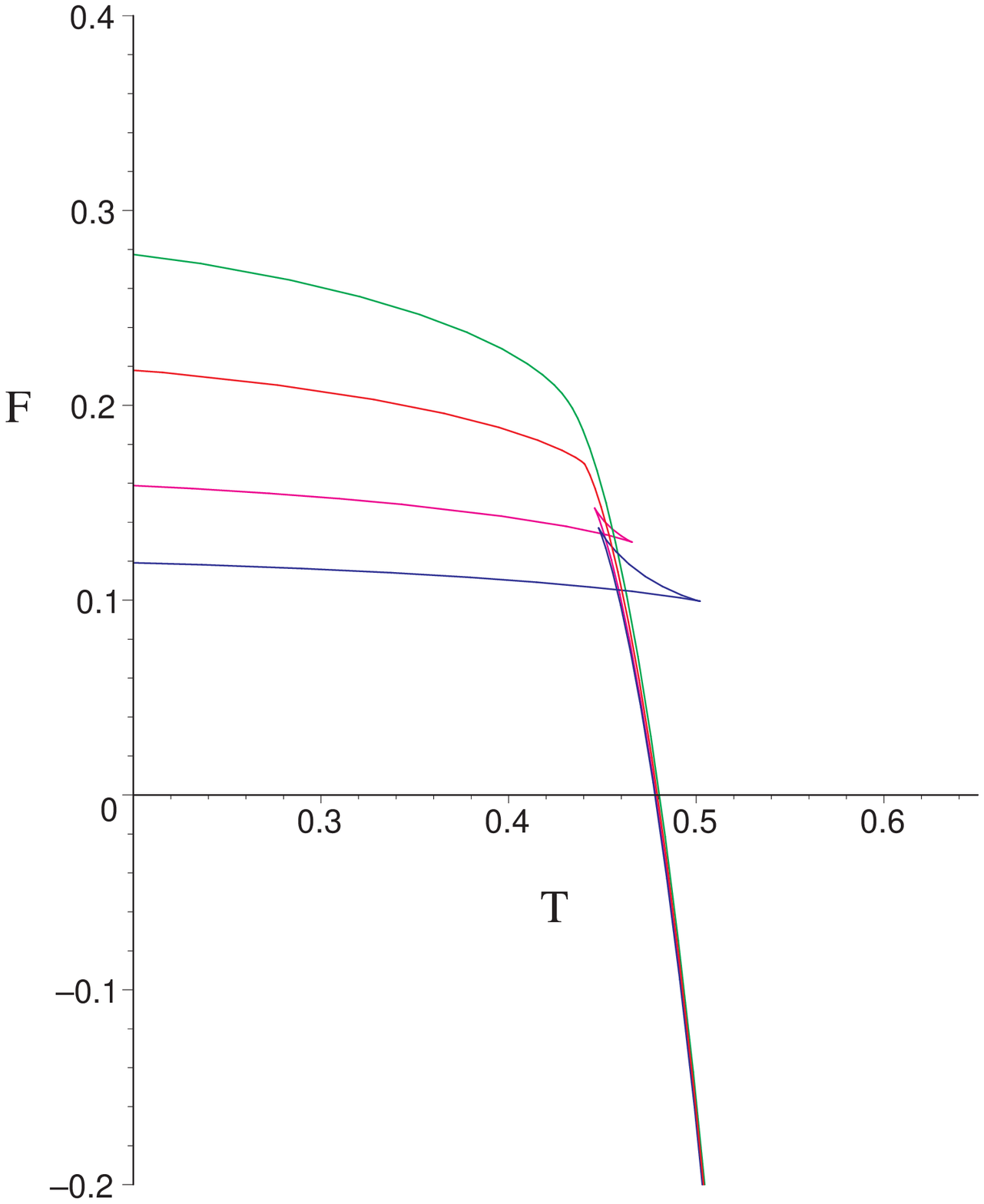}
\caption{Constant charge plots of the Helmholtz free energy $F$ vs. $T$. The blue and magenta colored curves have subcritical charges $q=0.12$ and $q=0.16$ respectively while the red colored curve is the critical curve $q=q_c=0.220$ and the green curve has $q=0.28>q_c$. }
\label{R7cs2ft}
\end{minipage}
\end{figure}

Fig.(\ref{R7cs2ph}) is a plot of the phase structure in the canonical ensemble, with the the heat capacity $C_q$ negative inside and positive outside its magenta colored curve of divergence. The green curve represents the zero of the Helmholtz energy $F$ which is negative to the right of the curve. Once again we may say that the thermal gas of R-charged particles has approximately zero free energy so that the black hole is globally stable against the thermal charged gas only to the right of the green curve. The constant charge lines which cross the $C_q$ stability curve two times will display a typical first order behaviour, with the stable small black hole branch separated from the unstable large black hole branch by an unstable branch. The constant charge line which is tangent to the $C_q$ stability curve is the critical curve while the point of tangency is the critical point of the second order phase transition. In terms of the thermodynamic variables temperature and charge it corresponds to $(T,q)=(0.440,0.220)$ while in terms of the black hole parameters it corresponds to  $(r_+,a)=(0.551,0.274)$.

In Fig.(\ref{R7cs2ft}) we show iso-charge plots of the free energy vs. temperature for some representative charges below and above the critical charge. The ``swallow tail'' shape of the subcritical plots indicates a first order phase behavior. It can also be seen that the critical temperature is less than the first order transition temperatures as in the single charge case.
The critical exponents can be obtained in a manner similar to the single charge case. They are  the same as the previous case and are given by

\begin{equation}
\label{R7cancrcs2}
{\alpha}=\frac{2}{3}, ~~{\beta}=\frac{1}{3},~~ {\gamma}=\frac{2}{3},~~ {\delta}=3 ~.
\end{equation}

We now turn to the phase behaviour in the grand canonical ensemble. The Gibbs free energy for the two charge case becomes
\begin{equation}
\label{Rcase2G}
G=\frac{1}{2}\,{r_+}^{2}-\frac{1}{2}\,{r_+}^{4}-a{r_+}^{2}-\frac{1}{2}\,{a}^{2}
\end{equation}

The heat capacity at constant potential $C_\phi$ becomes
\begin{equation}
\label{R7case2Cphi}
C_\phi=2 \pi\,{\frac {r_+ \left( 1+2\,{r_+}^{2}+2\,a \right) \, \left( {r_+}^{2}+a \right)  \left( {a}^{2}+4\,a{r_+}^{2}+3\,{r_+}^{4}+3\,{r_+}^{2} \right) }{2
\,{a}^{3}+{a}^{2}+6\,{a}^{2}{r_+}^{2}+6\,a{r_+}^{4}+2\,a{r_+}^{2}+{r_+}^{4}-{r_+
}^{2}+2\,{r_+}^{6}}}
\end{equation}
The susceptibilities $\kappa_T$ and $\alpha_\phi$ are given as

\begin{equation}
\label{R7case2beta}
\kappa_T={\frac { \left( 6\,{a}^{3}+14\,{a}^{2}{r_+}^{2}+7\,{a}^{2}+10\,a{r_+}^{4}+2\,a+8\,a{r_+}^{2}+{r_+}^{4}-{r_+}^{2}+2\,{r_+}
^{6} \right) }{2 \left( {r_+}^{2}+a \right)^{-1}\left(2\,{a}^{3}+{a}^{2}+6\,{a}^{2}{r_+}^{2}+6\,a{r_+}^{4}+2\,a{r_+
}^{2}+{r_+}^{4}-{r_+}^{2}+2\,{r_+}^{6}\right)}}
\end{equation}
and

\begin{equation}
\label{R7case2kappa}
\alpha_\phi=4\,{\frac {\sqrt {a}r_+ \left( {r_+}^{2}+a \right) ^{3/2} \left( 1+{r_+}^{2}+a \right) ^{3/2}\pi }{2\,{a}^{3}+{a}^{2}+6\,{a}^{2}{r_+}^{2}+6\,a{r_+}^{4
}+2\,a{r_+}^{2}+{r_+}^{4}-{r_+}^{2}+2\,{r_+}^{6}}}
\end{equation}

For the two charge case, as opposed to the single charge case, the Hessian alone is enough to determine the regions of stability in the parameter space, since the only polynomials which control the sign or the divergences of the susceptibilities are those which govern the divergence (as well as the sign) of the two heat capacities. In order to obtain the phase behaviour it will be useful to invert eq.(\ref{7cs1phi}) for $\phi$ and obtain $a$ as a function of $\phi$ and $r_+$,
\begin{equation}
\label{R7case2aphi}
a=-\frac{1}{2}-\frac{1}{2}\,{r_+}^{2}+\frac{1}{8}\,{\phi}^{2}+1/8\,\sqrt {{\phi}^{4}-8\,{\phi}^{2}+8\,{r_+}^{2}{\phi}^{2}+16+32\,{r_+}^{2}+16\,{r_+}^{4}}
\end{equation}

In terms of  $\phi$ and $r_+$ the temperature can be expressed as
\begin{equation}
\label{R7case2Tphi}
T={\frac {r_+ ( 4\,{r_+}^{2}+{\phi}^{2}+\sqrt {{\phi}^{4}-8\,{\phi}^{2}+8\,{r_+}^{2}{\phi}^{2}+16+32\,{r_+}^{2}+16\,{r_+}^{4}} ) }{\pi \,
 ( 4\,{r_+}^{2}-4+{\phi}^{2}+\sqrt {{\phi}^{4}-8\,{\phi}^{2}+8\,{r_+}
^{2}{\phi}^{2}+16+32\,{r_+}^{2}+16\,{r_+}^{4}} ) }}
\end{equation}

\begin{figure}[t!]
\begin{minipage}[b]{0.5\linewidth}
\centering
\includegraphics[width=3in,height=2.5in]{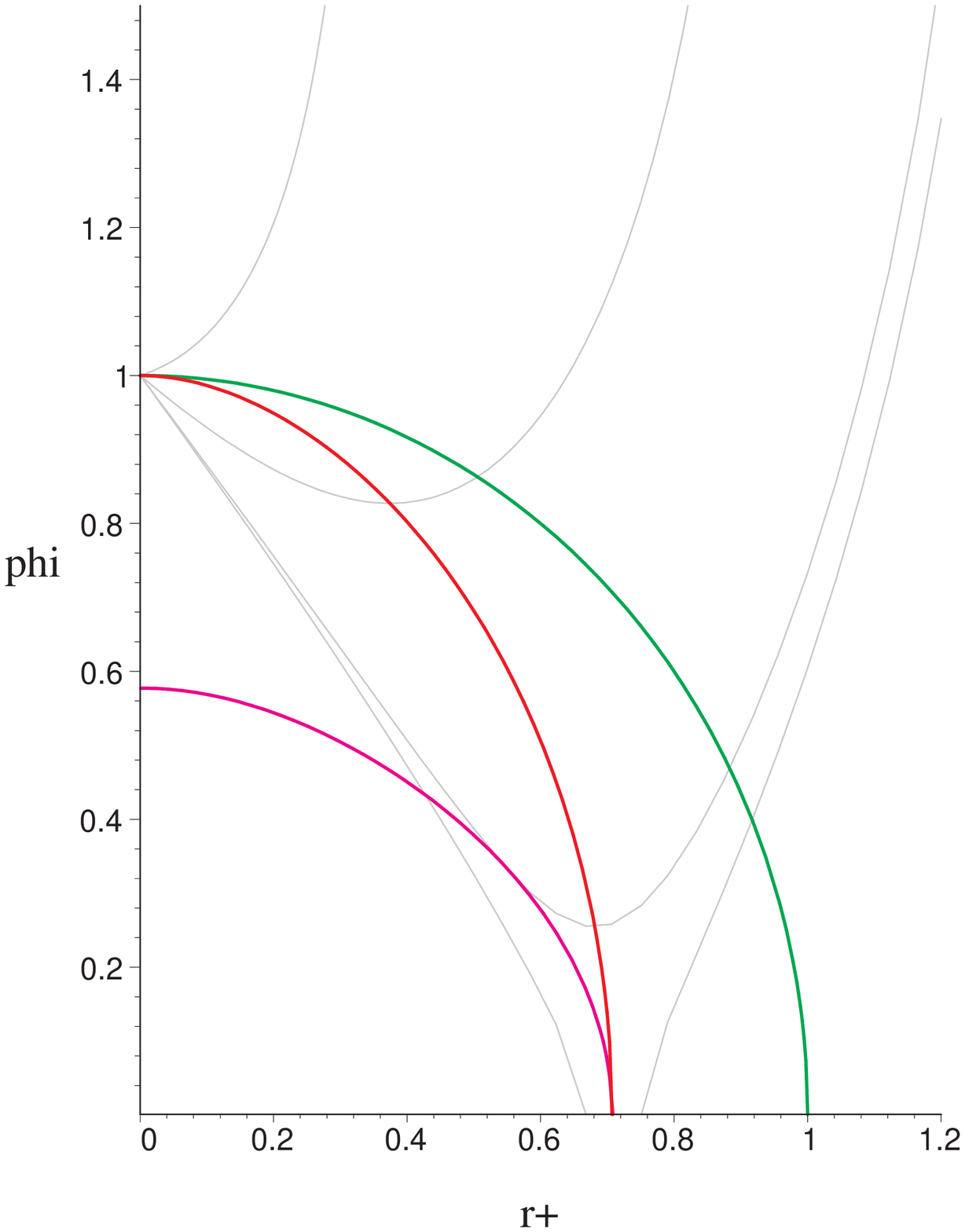}
\caption{Phase structure of the two charge case in the $\phi-r_+$ plane for the grand canonical ensemble. The stability curve showing infinities of $C_\phi$ is in red while the zeros of $\kappa_T$ and $G$ are in magenta and green respectively. The grey colored isotherms are at $T=0.120,0.320,0.440(T_c),0.450$ from top to bottom.}
\label{R7cs2phi1}
\end{minipage}
\hspace{0.6cm}
\begin{minipage}[b]{0.5\linewidth}
\centering
\includegraphics[width=3in,height=2.5in]{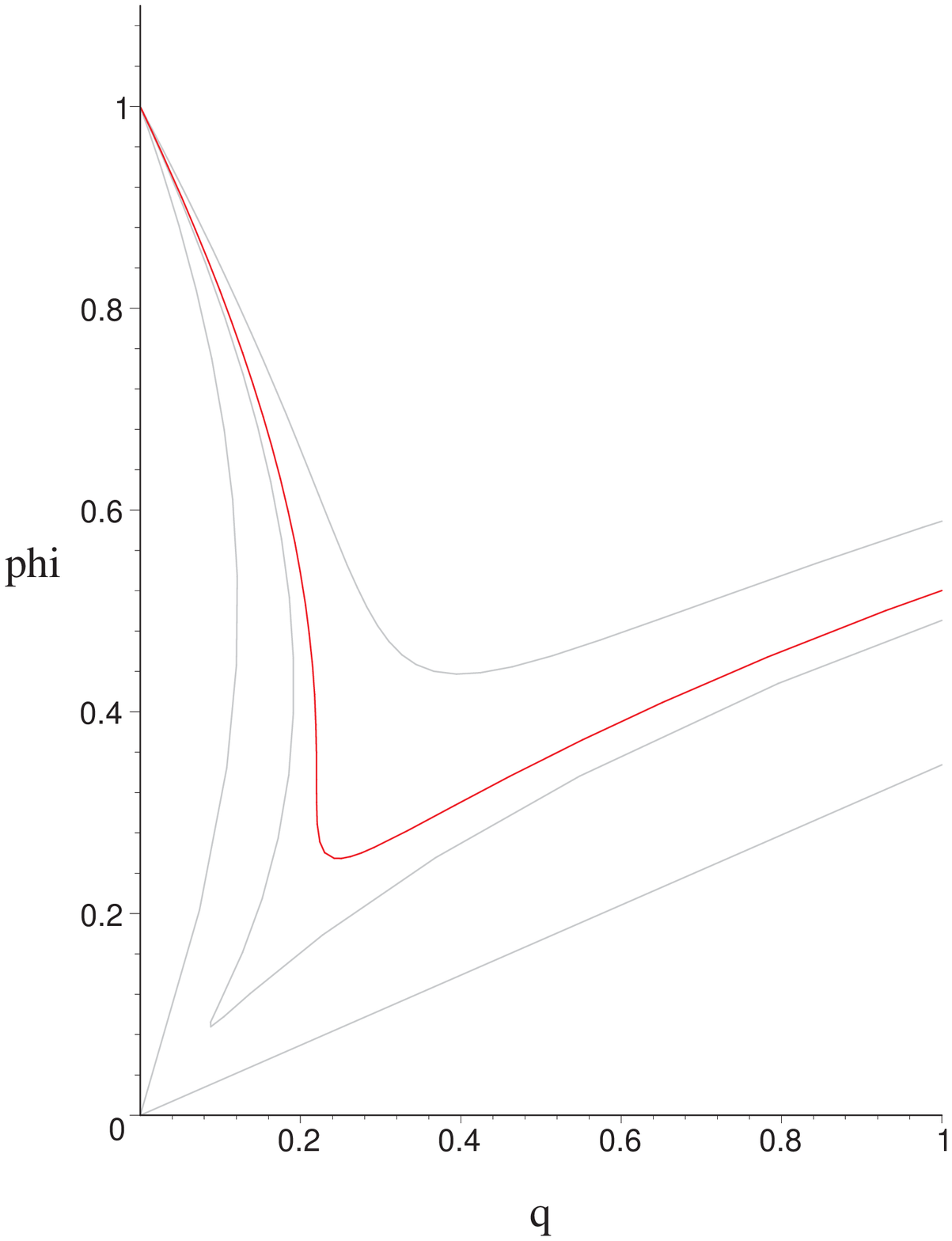}
\caption{Isotherms for the two charge case in the $\phi-q$ plane with $T=0.420,0.440(T_c),0.449,0.450$ from top to bottom. The critical isotherm is colored red while the rest are in grey. Here $\phi$ acts as the order parameter which jumps across a first order transition. }
\label{R7cs2iso}
\end{minipage}
\end{figure}

Using eq.(\ref{R7case2aphi}) the local and global stability conditions can be obtained with $\phi$ as a function of $r_+$ in a straightforward manner.
In fig.(\ref{R7cs2phi1}) we plot the grand canonical phase structure in the $\phi-r_+$ plane. The red colored local stability curve represents the divergence of $C_\phi$ which is negative inside the curve. The green colored curve is the Hawking-Page curve below which the black hole is globally unstable. From the figure it is apparent that the locally unstable regions are always globally unstable while there exists a region of meta-stability between the Hawking-Page curve and the $C_\phi$ curve. Note that the $C_\phi$-stability curve meets the $y$-axis at $\phi=1$ so that beyond this potential the black holes are globally as well as locally stable for all temperatures. Below $\phi=1$ the black hole exhibits a typical ``Davies phase '' behaviour as encountered in the case of the RN-AdS black holes in the grand canonical ensemble for example. The magenta colored curve inside the $C_\phi$-stability curve corresponds to the zeros of the susceptibility $\kappa_T$ so that it is positive inside the magenta curve, negative in between the magenta and the red curves and positive outside the red curve. The stability of $\kappa_T$ is important in the consideration of isotherms, some of which have been shown in the figure in grey color. Further, in fig.(\ref{R7cs2iso}) we plot isotherms in the $\phi-q$ plane in the vicinity of the critical isotherm corresponding to $T=T_c=0.440$. Here, just as in the case of the RN-AdS black holes in the canonical ensemble (see \cite{johnson1}), $\phi$ is the order parameter which jumps across a first order phase transition at constant $q$.

The thermodynamic curvature for the two charge case can be similarly found to be
\begin{equation}
\label{R7case2curv}
R=\frac{1}{2}\,\frac{AB\,\left(2\,{a}^{3}+{a}^{2}+6\,{a}^{2}{r_+}^{2}-3\,{r_+}^{2}+6\,a{r_+}^{4}-{r_+}^{4}+2\,{r_+}^{6}
\right)}{r_+\,C\,\left({2\,{a}^{3}+{a}^{2}+6\,{a}^{2}{r_+}^{2}+6\,a{r_+}^{4
}+2\,a{r_+}^{2}+{r_+}^{4}-{r_+}^{2}+2\,{r_+}^{6}}\right)^2}
\end{equation}
where the polynomials $A$, $B$, $C$ are given as
\begin{eqnarray}
A&=&2a^2+a+4ar_+^2+2r_+^2+2r_+^4\nonumber\\
B&=&6r_+^6+r_+^2+7r_+^4+18ar_+^4+14ar_+^2+7a^2+6a^3+18a^2r_+^2+2a\nonumber\\
C&=&3a^2+6ar_+^2+2a+3r_+^4+3r_+^2
\end{eqnarray}

\begin{figure}[t!]
\begin{minipage}[b]{0.5\linewidth}
\centering
\includegraphics[width=3in,height=2.5in]{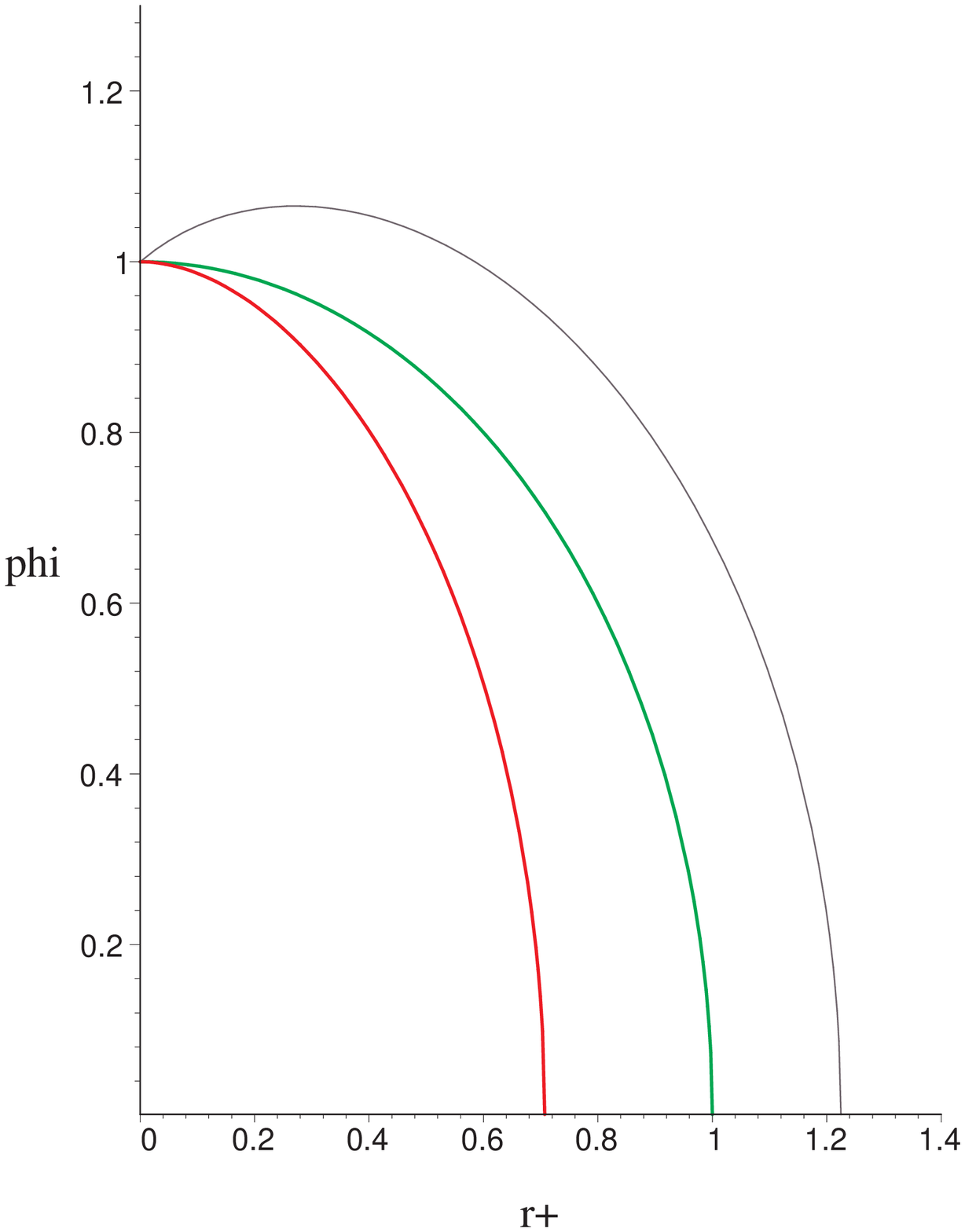}
\caption{Plot of zeros (in black) and infinities (in red) of $R$  in the $\phi-r_+$ plane for the two charge case. $R$ is positive outside the black curve. The green curve is the Hawking-Page curve. }
\label{R7cs2R}
\end{minipage}
\hspace{0.6cm}
\begin{minipage}[b]{0.5\linewidth}
\centering
\includegraphics[width=3in,height=2.5in]{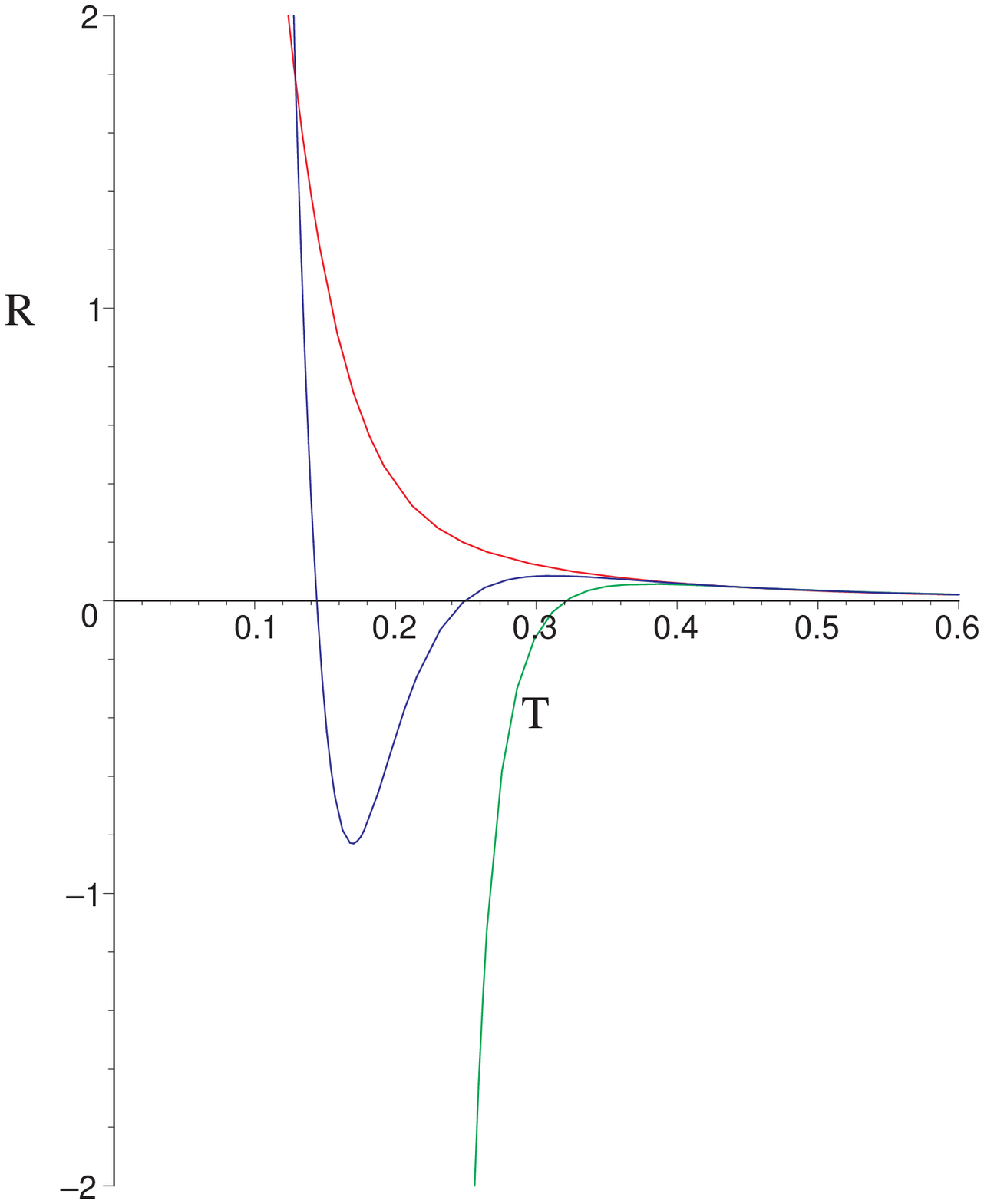}
\caption{Isopotential plots of $R$ vs. $T$ for $\phi=0.98, 1.05, 1.12$ in green, blue and red respectively. }
\label{R7cs2rt}
\end{minipage}
\end{figure}

The state space curvature diverges as the square of the heat capacity $C_{\phi}$ and also changes sign by passing through zero as dictated by the polynomial in the numerator of its expression in eq. (\ref{R7case2curv}). The zeros and divergences of the state space curvature have been represented in fig.(\ref{R7cs2R}). The curvature also shows a divergence at zero temperature for $\phi>1$. However, we shall ignore this divergence as thermodynamics breaks down  at extemality and quantum effects assume significance. It is apparent that the zeros of the curvature bear no relation to the Hawking-Page transition as was seen in the case of RNAdS black holes in four dimensions, \cite{tapo2}. Interestingly, it can be see from the figure that for $1.07>\phi>1$ the curvature turns negative briefly before becoming positive again while for $\phi>1.07$ it is positive for all temperatures. Fig.(\ref{R7cs2rt}) displays isopotential plots of $R \,vs. \,T$ for some representative values of the potential. The behaviour of $R$ at large temperatures is the same as in eq.(\ref{Rdecay1}), namely it decreases as the cube of inverse temperature.

\subsection{k=1, Case 3}

The case of three equal charges for the $D=5$ R-charged black holes has been discussed in detail in \cite{johnson1} and \cite{johnson2} where it was shown that these black holes are equivalent to the RN-AdS black holes in five dimensions. This is because on setting the three charges equal the Maxwell fields decouple from the scalar fields obtained by compactification on $S^5$. Let us now briefly discuss the thermodynamics and phase structure in this case. 

The first law for the three charge case becomes

\begin{equation}
\label{R7cs31stlaw}
dM=TdS+3\phi dq
\end{equation}

Just as in the two charge case we reinterpret the total charge $3q$ as the charge $q$ of the black hole. In this case there exist extremal black holes as can be seen from the expression for the temperature
\begin{equation}
\label{7cs2T}
T=\frac{1}{2}\,{\frac {{r_+}^{4}+2\,{r_+}^{6}+3\,a{r_+}^{4}-{a}^{3}}{\pi \, \left( {r_+}^{2}+a \right) ^{3/2}{r_+}^{2}}}
\end{equation}

The heat capacity $C_q$ and $C_\phi$ are given respectively as
\begin{equation}
\label{Rcs3cq}
C_q=6\,{\frac { \left( {{r_+}^{4}+2\,{r_+}^{6}+3\,a{r_+}^{4}-{a}^{3}} \right) \pi \, \left( {r_+}^{2}+a \right) ^{3/2}}{5\,{a}^{3}+12\,{a}^{2}{r_+}^{2}+
4\,a{r_+}^{2}+9\,a{r_+}^{4}+2\,{r_+}^{6}-{r_+}^{4}}}
\end{equation}
and
\begin{equation}
\label{Rcs3cphi}
C_\phi=6\,{\frac { \left({{r_+}^{4}+2\,{r_+}^{6}+3\,a{r_+}^{4}-{a}^{3}} \right) \pi \, \left( {r_+}^{2}+a \right) ^{3/2}}{{a}^{3}+4\,{a}^{2}{r_+}^{2}+5\,a
{r_+}^{4}+2\,{r_+}^{6}-{r_+}^{4}}}
\end{equation}
Similar to the two charge case, the susceptibilities give no additional information regarding the phase structure. 

The state space scalar curvature is also obtained in a straightforward manner as
\begin{eqnarray}
\label{R7cs3R}
R &=& \frac{r_+^2\left(3a^3 + 8a^2r_+^2 + 7ar_+^4 - 3r_+^4 + 2r_+^6\right)}{3\pi\left(a^3 + 4a^2r_+^2 + 5ar_+^4 + 2r_+^6 - r_+^4\right)^2\left(r_+^4 + 2r_+^6 + 3ar_+^4 - a^3\right)}
\nonumber\\
&\times& \left(6r_+^6 + r_+^4 + 15ar_+^4 + 4ar_+^2 + 12a^2r_+^2 + 3a^3\right)\sqrt{r_+^2 + a}
 \end{eqnarray}

\begin{figure}[t!]
\begin{minipage}[b]{0.5\linewidth}
\centering
\includegraphics[width=3in,height=2.5in]{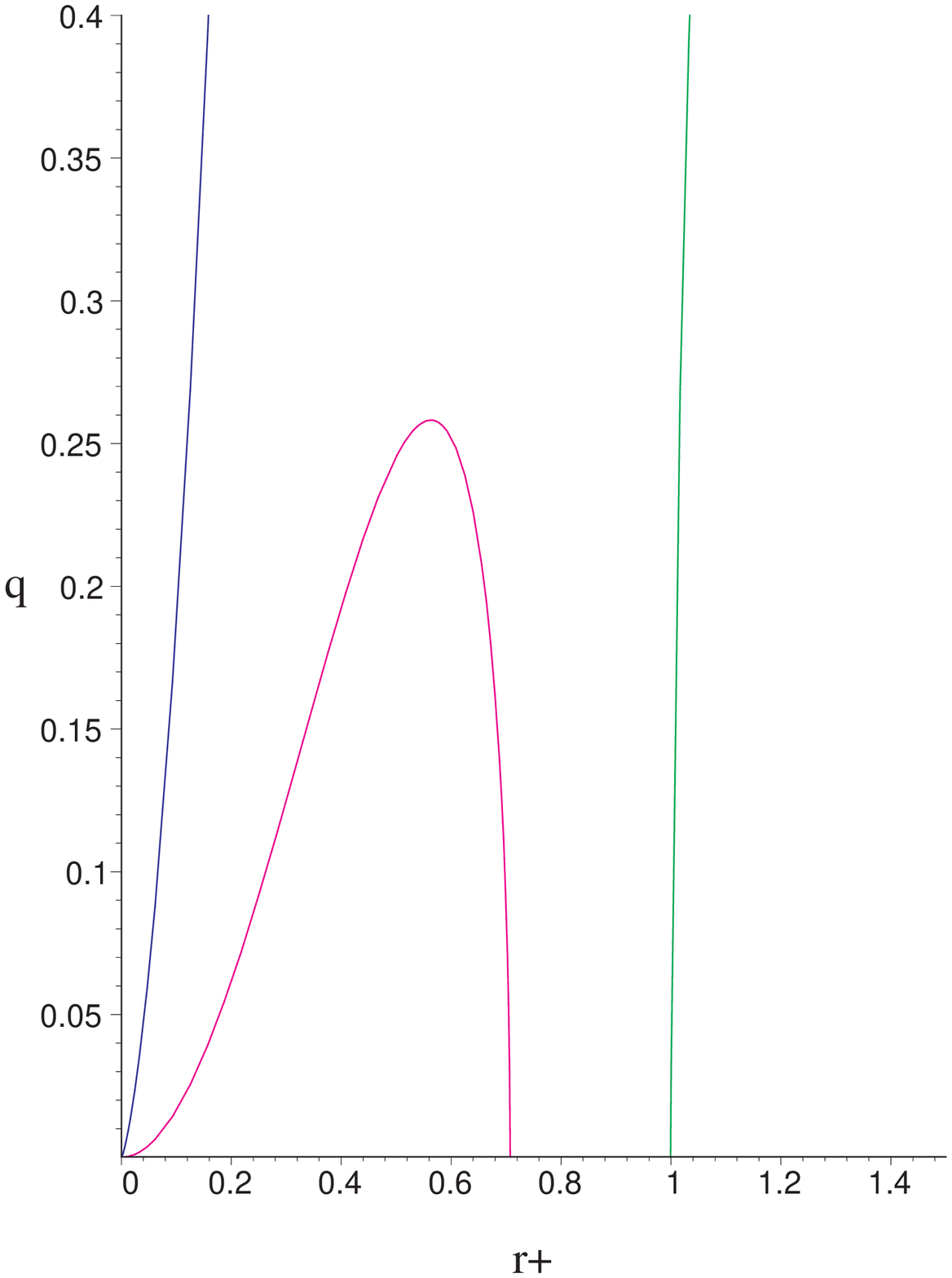}
\caption{$q-r_+$ plane plot of the phase structure of the canonical ensemble for the three charge case showing the infinities of $C_q$ in magenta, zeros of Helmholtz energy in green and the extremal curve in blue. The critical charge $q_c=0.258$.}
\label{R7cs3cn}
\end{minipage}
\hspace{0.6cm}
\begin{minipage}[b]{0.5\linewidth}
\centering
\includegraphics[width=3in,height=2.5in]{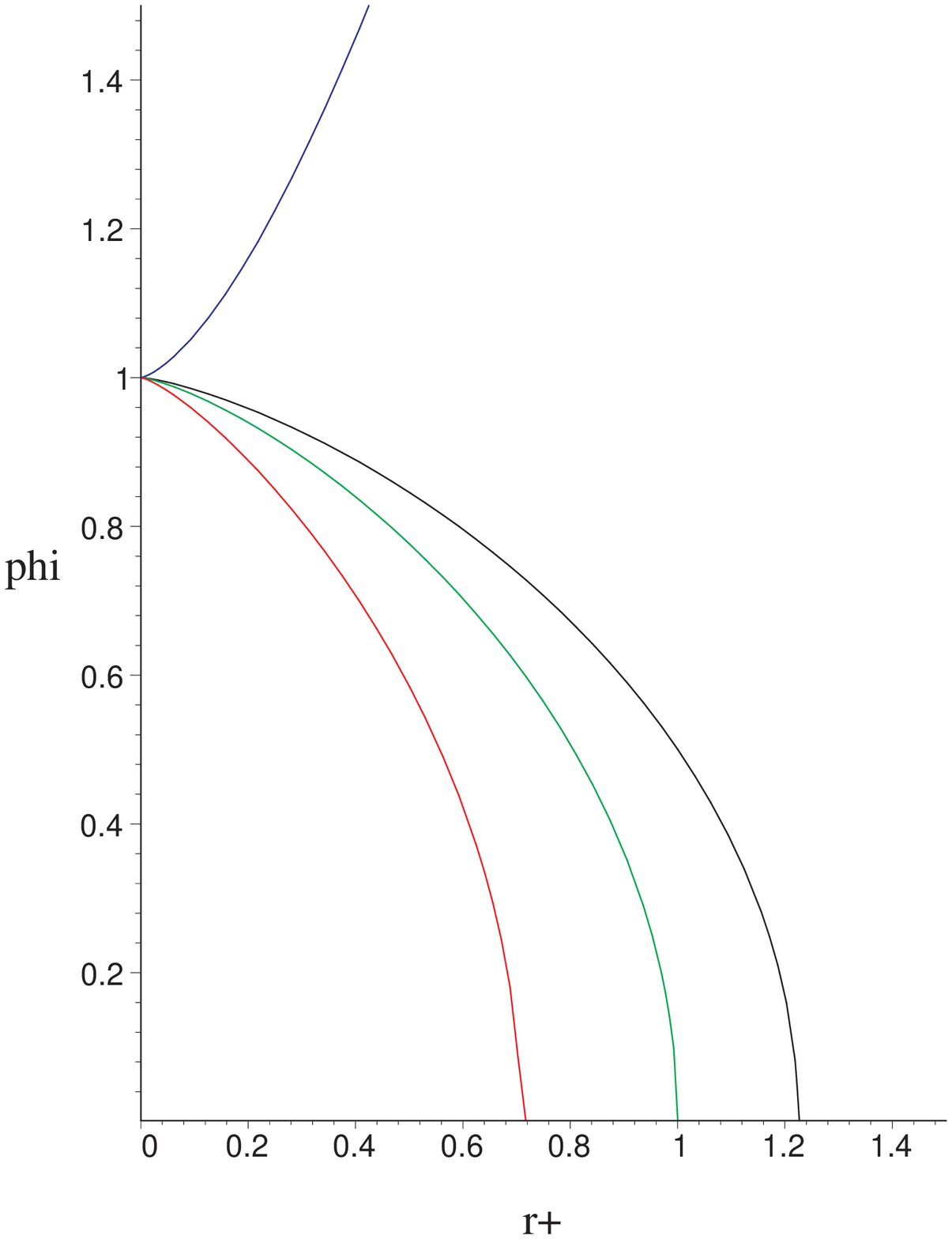}
\caption{$\phi-r_+$ plane plot of the phase structure in the grand canonical ensemble for the three charge case showing the infinities of $C_\phi$ in red, zeros of the Gibbs energy in green and the extremal curve in blue. Additionally, the zeros of $R$ are shown in black. }
\label{R7cs3gcn}
\end{minipage}
\end{figure}

In fig.(\ref{R7cs3cn}) we show the phase structure in the canonical ensemble. Constant charge processes exhibit a phase coexistence behaviour for $q<q_c=0.258$ while at the critical value of the charge the black hole undergoes a second order phase transition between the small black hole and the large black hole at $T_c=0.441$. The critical values of the black hole parameters are $(a_c,{r_+}_c)= (0.016,0.563)$. The critical exponents can be calculated in a manner similar to the previous cases and they are given as
\begin{equation}
\label{R7cancrcs3}
{\alpha}=2/3, ~~{\beta}=1/3,~~ {\gamma}= 2/3,~~ {\delta}=3 ~.
\end{equation}

This is expected, since these black holes are equivalent to the RNAdS black holes in five dimensions and the RNAdS black holes in all dimensions have the same set of critical exponents in the canonical ensemble as above, \cite{johnson2},\cite{wu}.

In fig.(\ref{R7cs3gcn}) we show the phase structure in the grand canonical ensemble. For $\phi<1$ we encounter the usual ``Davies'' phase behaviour followed by a Hawking-Page phase transition to a globally stable phase. For $\phi>1$ the black holes are locally as well as globally stable and also show extremality. Notice that now the zeros of $R$ do not exceed the $\phi=1$ line as in the two charge case so that for all values of potential greater than one the state space curvature remains positive.

\section{The case k=0}

R-charged black holes in $\mathcal{N}=8$ $D=5$ gauged supergravity with a planar horizon ( $k=0$) are equivalent to the near horizon geometry of near-extremal spinning D3-branes. The charges on the black holes are recovered by a Kaluza-Klein reduction of the ten-dimensional D3-brane solution on $S^5$, \cite{gub2}. The $k=0$ solution can also be recovered as large black hole limit of the compact $k=1$ solution. That is, one takes the limit $k\rightarrow 0^+$ by setting 
\begin{equation}
r_+\gg l ,~~ ~~a_i\gg l^2
\label{limit}
\end{equation}
where $l$ is the AdS length scale. The details  have been worked out extensively in \cite{gub2}.

Taking the limit as mentioned in the previous paragraph we write down the expression for the thermodynamic variables for the $k=0$ black hole. The mass and the charges are
\begin{eqnarray}
\label{R7k0M}
M=\frac{3}{2}\,{\frac { \left( {r_+}^{2}+a_{{1}} \right)  \left( {r_+}^{2}+a_{{2}} \right)  \left( {r_+}^{2}+a_{{3}} \right) }{{r_+}^{2}}}\nonumber\\
q_1={\frac {\sqrt {a_{{1}}}\sqrt { \left( {r_+}^{2}+a_{{1}} \right)  \left( {r_+}^{2}+a_{{2}} \right)  \left( {r_+}^{2}+a_{{3}}
 \right) }}{r_+}},~~{\rm etc.}
\label{R7k0q}
\end{eqnarray}
while the entropy remains the same. 
Similarly, the temperature and the conjugate potentials are given by 
\begin{equation}
\label{R7k0T}
T=\frac{1}{2}\,{\frac {2\,{r_+}^{6}+{r_+}^{4}a_{{3}}+{r_+}^{4}a_{{2}}+{r_+}^{4}a_{{1}}-a_{{1}}a_{{2}}a_{{3}}}{\pi r_+^2 \,\sqrt {{r_+}^{2}+a_{{1}}}\sqrt {{r_+}^{2}+a_
{{2}}}\sqrt {{r_+}^{2}+a_{{3}}}}}
\end{equation}
and
\begin{equation}
\label{R7k0phi}
\phi_1={\frac {\sqrt {{r_+}^{2}+a_{{3}}}\sqrt {{r_+}^{2}+a_{{2}}}\sqrt {a_{{1}}}}{r_+\sqrt {{r_+}^{2}+a_{{1}}}}},\,\,etc.
\end{equation}

In the following we shall briefly sketch the phase structure and thermodynamic geometry of the flat horizon R-charged black holes for the three cases mentioned in 
eq.(\ref{R7cases}). For detailed discussions on the phase structure of planar R-charged black holes, we refer to \cite{sudipta}.

We shall now briefly discuss the single charge case, \emph{i.e} the first case. For the single charge case the temperature becomes
\begin{equation}
\label{k0cas1T}
T=\frac{1}{2}\,{\frac {2\,{r_+}^{2}+a}{\sqrt {{r_+}^{2}+a}\pi }}
\end{equation}

The heat capacity $C_q$ is given as
\begin{equation}
\label{k0cas1Cq}
C_q=6\,{\frac { \left( 2\,{r_+}^{2}+a \right)  \left( {r_+}^{2}+a \right) ^{3/2}{r_+}^{2}}{5\,a{r_+}^{2}-{a}^{2}+2\,{r_+}^{4}}}
\end{equation}

Similarly, the heat capacity $C_\phi$ and the susceptibilities $\kappa_T$ and $\alpha_\phi$ become, respectively,
\begin{equation}
\label{k0cas1Cphi}
C_\phi=2\,{\frac { \left( 2\,{r_+}^{2}+a \right)  \left( a-3\,{r_+}^{2} \right) {r_+}^{2}}{\sqrt {{r_+}^{2}+a} \left( a-2\,{r_+}^{2} \right) }},
\end{equation}

\begin{equation}
\label{k0cas1kt}
\kappa_T={\frac {-5\,a{r_+}^{2}+{a}^{2}-2\,{r_+}^{4}}{ \left( a-2\,{r_+}^{2} \right)}},
\end{equation}
and

\begin{equation}
\label{k0cas1bphi}
\alpha_\phi=4\pi\,{\frac { \left( -{r_+}^{2}+a \right)\sqrt {a}r_+}{a-2\,{r_+}^{2}}}.
\end{equation}

The Gibbs free energy and the Hehlmontz free energy are given respectively as
\begin{equation}
\label{k0cas1gibb}
G=-\frac{1}{2}\,\left( {r_+}^{2}+a \right) {r_+}^{2}
\end{equation}
and
\begin{equation}
\label{k0cas1hehl}
F=\frac{1}{2}\,{r}^{2} \left( -{r}^{2}+a \right)
\end{equation}

Note the Gibbs free energy is always negative in this case, so that the black hole is always globally stable in the grand canonical ensemble.

The thermodynamic curvature is obtained as
\begin{equation}
\label{R7k0cs1R}
R={\frac {3\,a-2\,{r_+}^{2}}{ \left( a-2\,{r_+}^{2} \right)  \left( 2\,{r_+}^{2}+a \right) \pi \,\sqrt {{r_+}^{2}+a}}}
\end{equation}

Note that the above expression can be obtained from the corresponding one for the $k = 1$ case in eq. (\ref{singlechargeRcase1}), by taking the large black hole limit of
eq. (\ref{limit}). This is expected, because of the thermodynamic equivalence of black holes in the case $k=0$ and the large black holes for $k = 1$. \footnote{For all the charge configurations
considered in this paper, this will always be the case, i.e the equilibrium state space scalar curvature for the $k=0$ cases can be obtained by taking the large black hole limit
of the corresponding configurations for the $k=1$ cases. }

Just as in the previous cases the state space curvature diverges along with the heat capacity at fixed potential. 
The behaviour of the curvature with temperature is the same as in the compact horizon case, namely the state space curvature goes as the inverse third power of the temperature. However, a prominent difference with the compact black hole cases is that $R$ diverges as $C_\phi$ and not as the square of $C_\phi$ as in the previous cases. This can be understood by referring to
the discussion on fig. (\ref{fig00c}) where it was observed that for $\phi > 1$, one of the zeroes of $R$ is closely aligned to the polynomial that appears in its denominator. It can
be checked that by taking the large black hole limit of eq. (\ref{limit}), that these curves exactly coincide, leading to observation above.

The thermodynamic behaviour of the R-charged black holes for the two charge and three charge cases in case of flat horizon is completely regular. Let us briefly discuss the two cases.

For the two charge case the temperature becomes independent of the parameter $a$ and is simply given as
\begin{equation}
\label{k0cas2T}
T=\frac{r_+}{\pi}
\end{equation}

The Gibbs free energy remains negative while the Helmhontz free energy becomes
\begin{equation}
\label{k0cas2F}
F=\frac{1}{2}\, \left( {r_+}^{2}+a \right)  \left( -{r_+}^{2}+3\,a \right)
\end{equation}

The state space curvature is given as the simple expression
\begin{equation}
R=\frac{1}{2\pi}\,{\frac {1}{ \left( {r_+}^{2}+a \right) r_+}}
\end{equation}

For the three charge case the temperature becomes
\begin{equation}
\label{k0cas3T}
T=\frac{1}{2}\,{\frac {\sqrt {{r_+}^{2}+a} \left( -a+2\,{r_+}^{2} \right) }{\pi \,{r_+}^{2}}}
\end{equation}

The Gibbs free energy is once again always negative while the Helmhontz free energy becomes

\begin{equation}
\label{k0cas3F}
F=\frac{1}{2}\,{\frac { \left( {r_+}^{2}+a \right) ^{2} \left( -{r_+}^{2}+5\,a \right) }{\pi \,{r_+}^{2}}}
\end{equation}

The state space curvature becomes
\begin{equation}
R=-{\frac {{r_+}^{2} \left( 3\,a+2\,{r_+}^{2} \right) }{\pi \, \left( {r_+}^{2}+a \right) ^{3/2} \left( a-2\,{r_+}^{2} \right)  \left( 2\,{r_+}^{2}+a
 \right) }}
\end{equation}
It can be verified that for both the two charge and the three charge cases at high temperatures, the thermodynamic curvature goes as the inverse third power of the temperature, which 
therefore appears to be the case for all black holes in D=5.

\section{R-charged black holes in $AdS_4$}
Black holes in $N=8$ $D=4$ supergravity have four R-charges corresponding to the $SO(8)$ gauge symmetry arising out of isometries in $S^7$ of $AdS_4\times S^7$. These black holes are obtained by a Kaluza-Klein compactification of near horizon region of near extremal M2- branes. The thermodynamics and phase behaviour (mainly the single charge case) of these black holes have been studied in \cite{gub2} to which we refer for the general thermodynamic expressions. In this section we shall briefly discuss the additional features in the grand canonical ensemble for the single charge case and the scaling behaviour in the canonical ensemble for the three charge case of these black holes. By the latter we mean that three charges are set equal while the fourth charge is set to zero. The case of all four charges being equal corresponds to the RN-AdS black hole in four dimensions, \cite{johnson1}.

The mass, charge and entropy corresponding to the single charge black hole can be expressed in terms of the black hole radius $r_+$ and its charge parameter $a$ as
\begin{equation}
\label{M2mass}
M=2\,r_++2\,{r_+}^{3}+2\,{r_+}^{2}a+a,
\end{equation}

\begin{equation}
\label{M2charge}
q=\sqrt {a}\sqrt {r_++a}\sqrt {1+{r_+}^{2}},
\end{equation}
and

\begin{equation}
\label{M2entr}
S=4\,\pi \,{r_+}^{3/2}\sqrt {r_++a}.
\end{equation}

Just as in the single charge case for D=5 the BPS bound can be shown to be saturated only for $r_+=0$. The intensive variables temperature and potential are given as

\begin{equation}
\label{M2temp}
T=\frac{1}{4}\,{\frac {1+3\,{r_+}^{2}+2\,r_+a}{\pi \,\sqrt {r_+}\sqrt {r_++a}}},
\end{equation}
and

\begin{equation}
\label{M2pot}
\phi={\frac {\sqrt {a}\sqrt {1+{r_+}^{2}}}{\sqrt {r_++a}}}.
\end{equation}

\begin{figure}[t!]
\centering
\includegraphics[width=3in,height=2.5in]{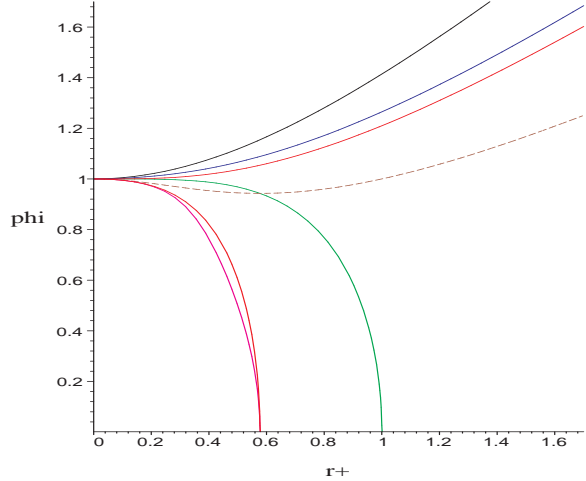}
\caption{ $\phi-r_+$ plot for the grand canonical ensemble of single R-charged black holes in D=4. The plot shows divergence and zeros of $C_\phi$ in red and blue respectively, zeros of $\kappa_T$ in magenta, zeros of $\alpha_\phi$ in dotted brown, zeros of the Gibbs potential in green and the physical limit curve $\phi=\sqrt{r_+^2+1}$ in black.}
\label{M2figgc}
\end{figure}

The heat capacities and the susceptibilities are given as
 
\begin{equation}
\label{M2cphi}
C_\phi=4\,{\frac { \left( 1+3\,{r_+}^{2}+2\,r_+a \right) \pi \,{r_+}^{3/2}\sqrt {r_++a} \left( r_+a-2-2\,{r_+}^{2} \right) }{-3\,{r_+}^{4}+2\,{a}^{2}{r_+}^{2}+1-3
\,r_+a-2\,{r_+}^{2}-a{r_+}^{3}}}
\end{equation}

\begin{equation}
\label{M2cq}
C_q=4\,{\frac { \left( 1+3\,{r_+}^{2}+2\,r_+a \right)  \left( 2\,{r_+}^{3}+2\,{r_+}^{2}a+2\,r_++3\,a \right) \pi \,{r_+}^{3/2}\sqrt {r_++a}}{-a+6\,a{r_+}^{4}+3
\,{r_+}^{5}+2\,r_+{a}^{2}-r_++7\,{r_+}^{2}a+2\,{r_+}^{3}}}
\end{equation}

\begin{equation}
\label{M2compr}
\kappa_T=-{\frac { \left( r_++a \right)  \left( -a+6\,a{r_+}^{4}+3\,{r_+}^{5}+2\,r_+{a}^{2}-r_++7\,{r_+}^{2}a+2\,{r_+}^{3} \right) }{r_+ \left( -3\,{r_+}^{4}+2\,{a}^{2
}{r_+}^{2}+1-3\,r_+a-2\,{r_+}^{2}-a{r_+}^{3} \right) }}
\end{equation}

\begin{equation}
\label{M2expan}
\alpha_\phi=4\,{\frac {\sqrt {a}\sqrt {1+{r_+}^{2}} \left( r_++a \right)  \left( -1-{r_+
}^{2}+2\,r_+a \right) \pi \,\sqrt {r_+}}{-3\,{r_+}^{4}+2\,{a}^{2}{r_+}^{2}+1-3
\,r_+a-2\,{r_+}^{2}-a{r_+}^{3}}}
\end{equation}

We re-emphasize that while the negative sign of the isothermal compressibility $\kappa_T$ implies a thermodynamic instability the negative sign of the ``isobaric'' expansivity $\alpha_\phi$ does not imply any such instability.
 
For the single charge case the canonical ensemble is simpler than the corresponding case for D=4 in that it does not exhibit a phase coexistence behaviour for any charge. The details of canonical ensemble phase behaviour have been worked out in \cite{gub2}. We shall now briefly discuss the phase behaviour in the grand canonical ensemble.

The Gibbs free energy for the single charge black holes is obtained as

\begin{equation}
\label{M2gibb}
G=-r_+ \left( {r_+}^{2}+r_+a-1 \right) 
\end{equation}

In order to express the thermodynamic quantities in terms of the potential we find it convenient to invert eq.(\ref{M2pot}) and express the parameter $a$ in terms of $r_+$ and $\phi$
\begin{equation}
a={\frac {r_+}{1-{\phi}^{2}+{r_+}^{2}}}
\end{equation}

Using the above substitution for $a$ every other thermodynamic variable may be expressed n terms of the potential and the horizon radius. In fig.(\ref{M2figgc}) we plot the grand canonical phase structure in the $\phi-r_+$ plane. The two red curves indicate the divergence in the heat capacity $C_\phi$ as well as the susceptibilities. The magenta curve within the lower red curve corresponds to the zeros of the isothermal capacitance $\alpha_T$ which is the same as the infinities of the heat capacity at constant charge $C_q$. Notice the difference with the D=5 case, fig.(\ref{R7c1ph1}), which has an additional branch of zeros of $\alpha_T$ for $\phi>1$. Incidentally, the brown curve corresponding to the zeros of $\kappa_T$ intersects the green colored Hawking-Page curve at its minima at $r_+=1/\sqrt{3}$ and $\phi=0.943$. The phase behaviour can be described as follows. The heat capacity $C_\phi$ is positive between the lower and upper red curves and between the black and the blue curves. The isothermal capacitance $\alpha_T$ is positive below the magenta curve and between the two red curves. The ``expansivity'' $\kappa_T$ is positive between the dotted brown curve and the lower red curve and between the upper red curve and the black curve. Moreover, for $\phi<0.943$ the globally stable black holes always have positive expansivity while for $\phi>0.943$ the globally stable black holes have a negative expansivity up to a certain horizon radius determined by the intersection of the constant potential straight line with the brown curve. The region of full thermodynamical stability now only corresponds to the one in between the two red curves. The region in between the blue curve and the black curve, even though thermally stable, is electrically unstable since there is no upper branch of the magenta curve as in fig.(\ref{R7c1ph1}) to effect a sign change in the compressibility. As usual, the region between the lower red curve and the green Hawking-Page curve is metastable.

The thermodynamic curvature for the single charge case is given as 
\begin{eqnarray}
\label{RD4curv}
R &=&
{\frac {\left( -a+2\,a{r_+}^{4}+6\,{r_+}^{5}+2\,r_+{a}^{2}+7\,{r_+}^{2}a+6\,{r_+}^{3}-4\,{a}^{2}{r_+}^{3} \right)}
{\left( 3\,{r_+}^{4}-2\,{r_+}^{2}{a}^{2}-1+3\,r_+a+2\,{r_+}^{2}+{r_+}^{
3}a \right) ^{2}}}\nonumber\\
&\times& \frac{A\left( -2\,{r_+}^{2}{a}^{2}-1+r_+a-{r_+}^{
3}a+{r_+}^{4} \right)}{\left( 2\,{r_+}^{3}+2\,{r_+}^{2}a+2\,r_++3\,a\right)}
\end{eqnarray}
where
\begin{equation}
A=\frac{9}{4}{\frac { \left( 1+{r_+}^{2} \right) \sqrt{r_+ + a}}{\sqrt {r_+} \left( 1+3\,{r_+}^{2}+2\,r_+a \right) \pi }}
\end{equation}
The curvature diverges in an expected manner as the square of the heat capacity $C_\phi$ or the susceptibilities. The zeros of $R$ have a similar distribution in the $\phi-r_+$ plane as the zeros of $R$ for the single charge case of R-charges black holes in D=5. This means that one of the polynomial in the numerator of the expression for the curvature has its zeros which closely follow the upper red branch in fig.(\ref{M2figgc}). Further, it can be verified that at large temperatures the state space curvature goes as
\begin{equation}
\label{D4RT}
R\sim \frac{1}{T^2}
\end{equation}

The thermodynamics of the k=0 case, which corresponds to a Kaluza-Klein compactification of the M2-brane, can be recovered from the k=1 case by taking the limit
\begin{equation}
r_+\gg l ~~~\mbox{and}~~~ a_i\gg l,
\label{limit1}
\end{equation}

The thermodynamic curvature for the single charge black hole with k=0 becomes,

\begin{equation}
\label{M5k0R}
R=\frac{9}{4}\,{\frac {2\,a-r_+}{\sqrt {r_++a} \left( 3\,r_++2\,a \right) \pi \,\sqrt {r_+} \left( 2\,a-3\,r_+ \right) }}
\end{equation}
Once again this can be understood by taking the large black hole limit, eq. (\ref{limit1}), of the thermodynamic curvature for the k=1 case.

For the three charge case, namely, three charges are set equal while the fourth charge is set to zero, the D=4 black hole exhibits a phase coexistence behaviour terminating in a critical point for the canonical ensemble, much like the two charge case for the D=5 black hole. The critical parameters are $(a,r_+,T,q)=(0.091,0.320,0.257,0.629)$ while the critical exponents are the same as in the D=5 black holes.
It can be verified that the thermodynamic curvature behaves in the same manner at high temperatures, as in eq.(\ref{D4RT}), for all D=4 black
holes.

\section{R-charged black holes in $AdS_7$}

R-charged black holes in  N=4, D=7 supergravity are obtained in the near horizon limit of the near extremal M5-branes. They carry two independent R-charges corresponding
 to the isometries of $S^4$ in the near horizon geometry $AdS_7\times S^4$. The thermodynamics and phase structure of the single charge D=7 black holes was first discussed 
in \cite{gub1}.
For the single charge compact horizon black holes, the mass, charge and entropy become

\begin{equation}
\label{M5mass}
M=\frac{5}{4}\,{r_+}^{4}+\frac{5}{4}\,{r_+}^{6}+\frac{5}{4}\,{r_+}^{2}a+a
\end{equation}

\begin{equation}
\label{M5charge}
q=\sqrt {a}\sqrt {{r_+}^{4}+a}\sqrt {{r_+}^{2}+1},
\end{equation}
and

\begin{equation}
\label{M5entr}
S=\pi \,{r_+}^{3}\sqrt {{r_+}^{4}+a}.
\end{equation}

while the intensive variables temperature and potential are

\begin{equation}
\label{M5temp}
T=1/2\,{\frac {2\,{r_+}^{2}+3\,{r_+}^{4}+a}{\pi \,\sqrt {{r_+}^{4}+a}r_+}},
\end{equation}
and

\begin{equation}
\label{M5pot}
\phi={\frac {\sqrt {a}}{\sqrt {{r_+}^{4}+a}}}.\end{equation}

The BPS bound is saturated at the naked singularity $r_+=0$ just as in the single charge cases of the previous black holes. The heat capacities and the susceptibilities are given 
as
\begin{equation}
\label{M5cphi}
C_\phi={\frac { \left( 2{r_+}^{2}+3{r_+}^{4}+a \right) \pi \sqrt
{{r_+}^{4}+a}{r_+}^{3} \left( a-5{r_+}^{4}-5{r_+}^{2} \right)
}{2{r_+}^{4}-{r_+}^{6}-3
{r_+}^{8}-2{r_+}^{4}a-3{r_+}^{2}a+{a}^{2}}}
\end{equation}

\begin{equation}
\label{M5cq}
C_q=-{\frac {{r_+}^{3} \left( 2{r_+}^{2}+3{r_+}^{4}+a \right)  \left(
5{r_+}^{2}a+6a+5{r_+}^{6}+5{r_+}^{4} \right) \pi \sqrt {{r_+}^{4}+a}}{2{
r_+}^{6}-{r}^{8}-3{r_+}^{10}-12{r_+}^{6}a+2{a}^{2}-17{r_+}^{4}a+3{r_+}
^{2}{a}^{2}-4{r_+}^{2}a}}
\end{equation}

\begin{equation}
\label{M5compr}
\kappa_T={\frac { \left( {r_+}^{4}+a \right)  \left(
-2{r_+}^{6}+{r_+}^{8}+3{r_+}^{10}+12{r_+}^{6}a-2{a}^{2}+17{r_+}^{4}a-3{r_+}^{2}{a}^{2}+4{r_+}^{2}a
 \right) }{{r_+}^{2} \left( -2{r_+}^{4}+{r_+}^{6}+3{r_+}^{8}+2{r_+}^{4}a+3
{r_+}^{2}a-{a}^{2} \right) }}
\end{equation}

\begin{equation}
\label{M5expan}
\alpha_\phi=4\,{\frac {\sqrt {a}\sqrt {{r_+}^{2}+1}r_+ \left( {r_+}^{4}+a
\right)  \left( 2{r_+}^{4}+2{r_+}^{2}-a \right) \pi
}{-2{r_+}^{4}+{r_+}^{6}+3{
r_+}^{8}+2{r_+}^{4}a+3{r_+}^{2}a-{a}^{2}}}
\end{equation}

The Helmholtz and the Gibbs free energies are obtained as

\begin{equation}
\label{M5hel}
F=\frac{1}{4}{r_+}^{4}-\frac{1}{4}{r}^{6}+\frac{3}{4}{r_+}^{2}a+a
\end{equation}
and
\begin{equation}
\label{M5gibb}
G=-\frac{1}{4}{r_+}^{2} \left( {r_+}^{4}-{r_+}^{2}+a \right)
\end{equation}

\begin{figure}[t!]
\begin{minipage}[b]{0.5\linewidth}
\centering
\includegraphics[width=3in,height=2.5in]{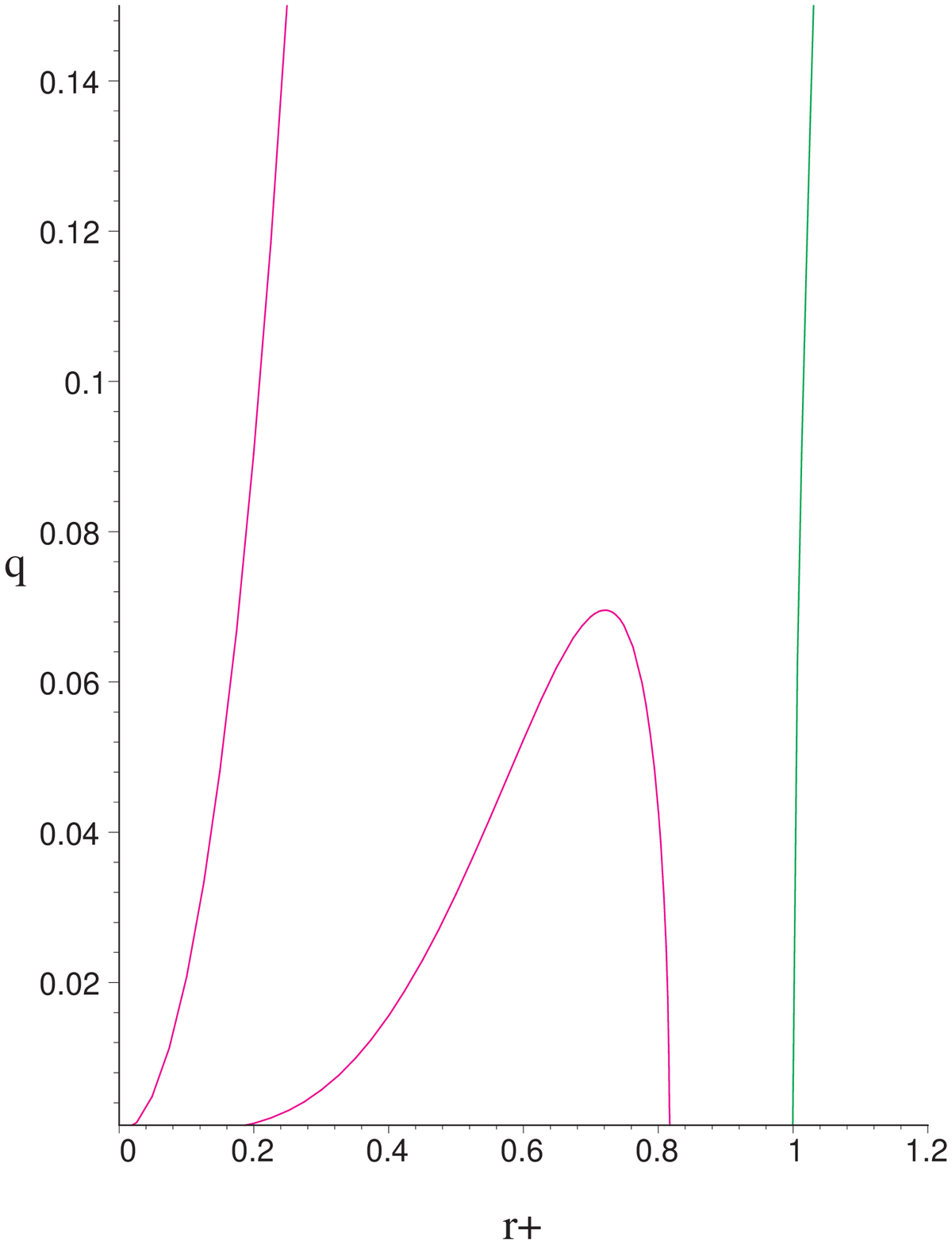}
\caption{$q-r_+$ plane plot of the phase structure in the canonical ensemble for the single charge black hole in D=7 supergravity. The magenta curves represent the infinities of $C_q$ and the green curve shows the zeros of $F$.}
\label{m5can}
\end{minipage}
\hspace{0.6cm}
\begin{minipage}[b]{0.5\linewidth}
\centering
\includegraphics[width=3in,height=2.5in]{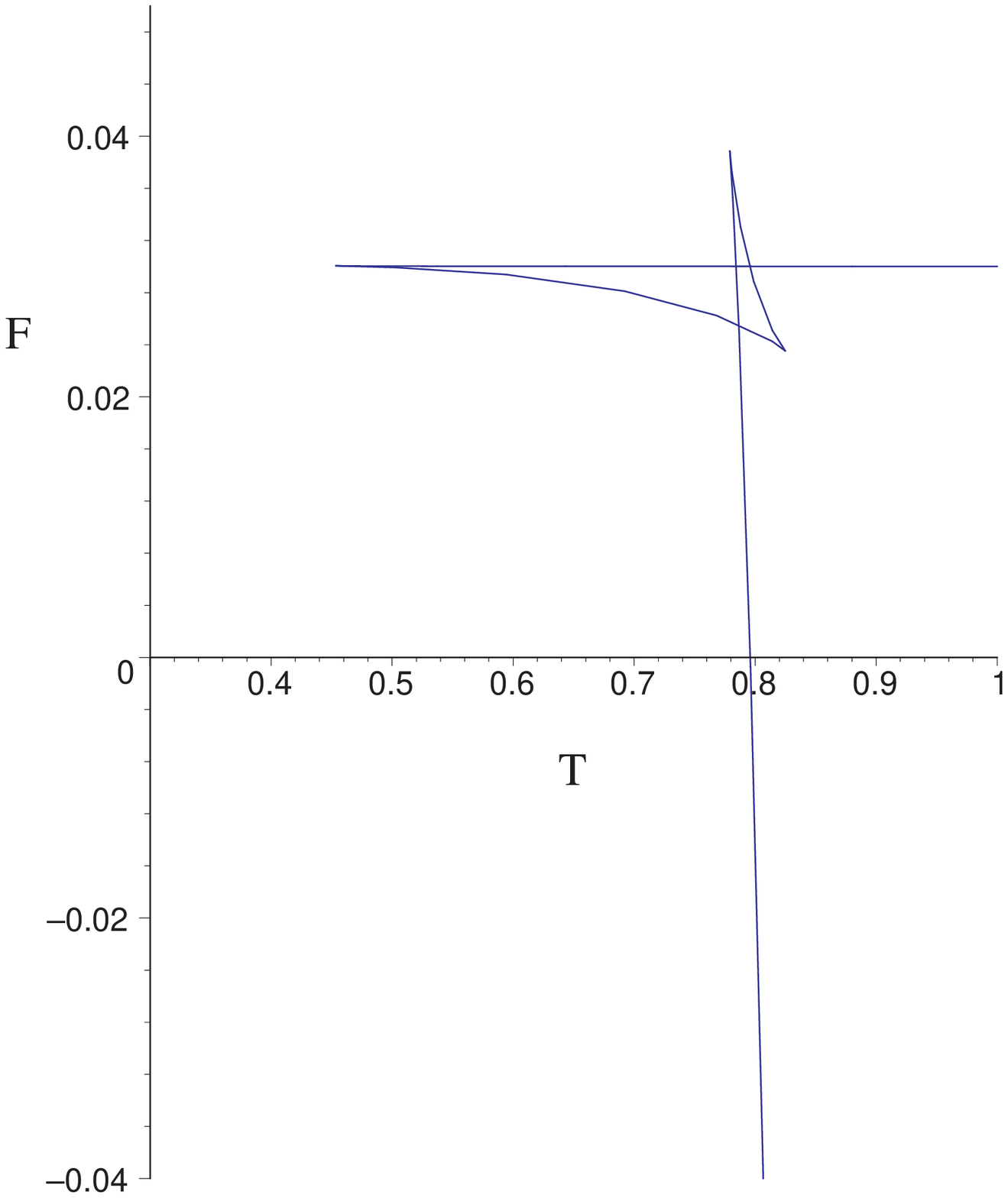}
\caption{Plot of the Helmholtz free energy vs temperature with the charge fixed at $q=0.03$.  }
\label{m5ft}
\end{minipage}
\end{figure}

In fig.(\ref{m5can}) we obtain the phase structure in the canonical ensemble in the $q-r_+$ plane. The two magenta curves, the first one being open and the second one being closed, indicate the divergence in the heat capacity $C_q$ while the green line indicates the zero of the Helmholtz free energy. The heat capacity $C_q$ is negative to the left of the first magenta curve and inside the second magenta curve while it is positive everywhere else. The free energy is negative to the right of the green curve. Because of the peculiar arrangement of the $C_q$-stability curves an unstable black hole branch exists for all temperatures starting from the ``Davies'' temperature while the stable branch itself displays a phase coexistence behaviour somewhere above the Davies temperature. In fig.(\ref{m5ft}) we further illustrate this behaviour in an isocharge plot of the Helmholtz free energy vs. temperature. The curve on the positive y-axis which is almost a straight line and moves rightwards towards infinity represents the unstable region to the left of the first magenta curve in fig.(\ref{m5can}) while the rest of the phase coexistence behavior is due to the crossing by the iso-charge lines of the lower magenta curve in fig.(\ref{m5can}). The critical point corresponds to
the parameters $(a,r_+,T,q)=(0.011,0.721,0.774,0.069)$. It can be verified that the critical exponents are the same as those for the previous black hole cases, namely,

\begin{equation}
{\alpha}=2/3, ~~{\beta}=1/3,~~ {\gamma}= 2/3,~~ {\delta}=3 ~.
\end{equation}

The thermodynamic curvature is given as

\begin{eqnarray}
R&=&
\frac{\left( -6\,{r_+}^{6}-21\,{r_+}^{8}-15\,{r_+}^{10}-4\,{r_+}^{2}a-17\,{r_+}^{4}a-10\,{r_+}^{6}a+5\,{r_+}^{2}{a}^{2}+2
\,{a}^{2} \right)}{ \left( 2\,{r_+}^{4}-{r_+}^{6}-3\,{r_+}^{8}-3\,{r_+}^{2}a-2\,{r_+}
^{4}a+{a}^{2} \right) ^{2}}\nonumber\\
&\times&
\frac {9 \left( 1+{r_+}^{2} \right) \left( {a}^{2}-{r_+}^{2}a+2\,{r_+}^{4}+{r_+}^{6}-{r_+}^{8}
 \right) \sqrt {{r_+}^{4}+a}}{\pi r \left( 5\,{r_+}^{6}+5\,{r_+}^{2}a+5\,{r_+}^{4}
+6\,a \right) \left( 2\,{r}^{2}+3\,{r}^{4}+a \right)}
\label{M2Rk1}
\end{eqnarray}

It can be verified that at large temperatures it behaves as
\begin{equation}
R\sim \frac{1}{T^5}
\end{equation}

for the k=0 case the state space curvature becomes
\begin{equation}
R=9\,{\frac { \left( a-{r_+}^{4} \right) r_+}{ \left( -3\,{r_+}^{4}+a
\right)  \left( 3\,{r_+}^{4}+a \right) \pi \,\sqrt {{r_+}^{4}+a}}}
\label{M2Rk0}
\end{equation}
Once again this can be recovered from eq.(\ref{M2Rk1}) by taking the
large black hole limit which for D=7 black holes corresponds to
$r\gg l,\,,a\gg l^4$.
It can be verified that for all black holes in $D = 7$, the equilibrium state space scalar curvature at high temperature behaves as  $R \sim \frac{1}{T^5}$. 

\section{Discussions and Conclusions}

The present study is a logical culmination of the investigation initiated in our previous papers \cite{tapo1} and \cite{tapo2}, where  we studied the thermodynamic geometry and
critical phenomena of asymptotically AdS black hoes in four dimensional Einstein Maxwell theory. In the present paper, we have elucidated the
issues of thermodynamic stability and the equilibrium state space geometry of R-charged black holes arising in gauged supergravity theories. Although the issue of
stability for these black holes have been extensively studied over the last few years, our work demonstrates certain novel features which complement earlier results on the
subject. 

For the case of compact R-charged black holes in $D=5$, $4$ and $7$, our results indicate a novel liquid-gas like first order phase transition in the canonical
ensemble, culminating at a second order critical point. 
Interestingly, we have shown that the critical exponents are identical for all the cases mentioned above. 
Further, they also turn out to be the same for those of the Kerr-AdS and KN-AdS black holes in $D=4$  \cite{tapo2} 
and the RN-AdS black holes in arbitrary dimensions \cite{johnson1}. This naturally suggests an interesting universality
in the  critical behaviour of asymptotically AdS black holes. 
We have further shown that in the grand canonical ensemble, the regions of stability for the single charged case are more constrained than the results obtained through
the positivity of the Hessian for the appropriate system. 

The equilibrium state space geometry for different charge configurations for the black holes mentioned above have also been considered in this paper and the corresponding 
state space scalar curvatures have been obtained.  The curvatures diverge at the Davies points, 
which correspond to the singularities in the heat capacity at constant potential in the grand canonical ensemble. Note that thermodynamic geometry
requires fluctuations in at least two extensive variables, and therefore naturally alludes to a grand canonical ensemble. 
Further, for the large black hole limit of the $k=1$ case, the state space scalar curvature has been shown to asymptote to the $k \to 0^+$ case, as expected. 

It was shown in \cite{caisoh} and \cite{gub2} that the divergence of the heat capacity $C_{\phi}$ in the grand canonical ensemble in the single
charge case signals a second order phase transition in the dual field theory. The critical exponents for this phase transition have been calculated in \cite{caisoh}. We point 
out here that the thermodynamic curvature does not scale as expected (see, e.g \cite{tapo2}) at these critical points (the Davies points). 
The reason for this is that at the Davies points, the system changes from an unstable to a stable branch.

We have also established the asymptotic behaviour of the scalar curvature at high temperatures. It has been shown that 
$R \sim \frac{1}{T^3}$ for the $D=5$ case, whereas for $D=4$ and $7$, it goes as $\frac{1}{T^2}$
and $\frac{1}{T^5}$ respectively. 
The nature of the scalar curvature is clear for conventional thermodynamic systems, as a correlation volume. For asymptotically flat black holes,
the interpretation is not quite well established. However, as a consequence of the gauge/gravity duality, the thermodynamics of
AdS black holes have a clear interpretation on the dual field theory side. 
Hence, it is expected that the scalar curvature and its asymptotic forms obtained in this paper should have specific interpretations in the
the strongly coupled dual field theory. We leave these issues for a future investigation.

\end{document}